\documentclass[%
twocolumn,
 amsmath,amssymb,
 aps,
pra,
]{revtex4-1}

\usepackage{graphicx}
\usepackage{dcolumn}
\usepackage{bm}

\begin{document}

\title{Optical spectroscopy of a microsized Rb vapour sample in magnetic fields up to 58 tesla}

\author{D. Ciampini\footnote{E-mail address: donatella.ciampini@unipi.it\\}}
\affiliation{Dipartimento di Fisica ``E. Fermi'', Universit\`a di Pisa, Largo B. Pontecorvo 3, 56127 Pisa, Italy}
\affiliation{INO-CNR, Via G. Moruzzi 1, 56124 Pisa, Italy}
\affiliation{CNISM, UdR Dipartimento di Fisica ``E. Fermi'', Universit\`a di Pisa, Largo B. Pontecorvo 3, 56127 Pisa, Italy}
\author{R. Battesti}
\affiliation{Laboratoire
National des Champs Magn\'etiques Intenses (UPR 3228,
CNRS-UPS-UGA-INSA), F-31400 Toulouse Cedex, France}
\author{C. Rizzo}
\affiliation{Laboratoire
National des Champs Magn\'etiques Intenses (UPR 3228,
CNRS-UPS-UGA-INSA), F-31400 Toulouse Cedex, France}
\author{E. Arimondo}
\affiliation{Laboratoire
National des Champs Magn\'etiques Intenses (UPR 3228,
CNRS-UPS-UGA-INSA), F-31400 Toulouse Cedex, France}
\affiliation{Dipartimento di Fisica ``E. Fermi'', Universit\`a di Pisa, Largo B. Pontecorvo 3, 56127 Pisa, Italy}
\affiliation{INO-CNR, Via G. Moruzzi 1, 56124 Pisa, Italy}

\date{\today}

\begin{abstract}
We use  a magnetometer probe based on the Zeeman shift of the rubidium resonant optical transition to explore the atomic magnetic response for a wide range of field values. We record optical spectra for fields from few tesla up to  60 tesla, the limit of the coil producing the magnetic field.  The atomic absorption is detected by the fluorescence emissions from a very small region with a submillimiter  size. We investigate a wide range of magnetic interactions  from the hyperfine Paschen-Back regime to the  fine one,  and the transitions between them.  The magnetic field measurement is based on the rubidium absorption itself. The rubidium spectroscopic constants were previously measured with high precision, except the excited state Land\'e $g$-factor that we derive from the position of the absorption lines in the  transition to the fine Paschen-Back regime. Our spectroscopic investigation, even if limited by the   Doppler broadening of the absorption lines, measures the field with a 20 ppm uncertainty at the explored high magnetic fields. Its  accuracy is limited to 75 ppm by the excited state Land\'e $g$-factor determination   \end{abstract}

\maketitle

\section{Introduction}
The present large effort of the quantum control research is the miniaturization and manipulation from the micron scale down to  the single atom. This objective is important for a complete quantum control and also for the development of new tools for applications, as geophysics, biophysics, brain imaging, and more. Recently a large attention  was concentrated on the magnetic response and the measurement of weak magnetic fields with a high spatial resolution.  Technologies capable of micron-scale magnetic microscopy include SQUID devices, scanning Hall probe microscopes, magnetic force microscopes, magneto-optical imaging techniques.  Magnetic fields may also be measured by detecting the Zeeman splitting for warm and ultracold atoms~\cite{KominisRomalis:2003,VengalattoreStamperKurn:2007},  nuclei in a ferromagnetic material~\cite{MaminRugar:2007}, or impurities in diamond (NV-centers)~\cite{TaylorLukin:2008,PhamWalsorth:2011,RondinJacques:2014}.\\
\indent The same large attention was not reserved to the measurement of high magnetic fields, whose application range is steadily growing.  Nowadays accurate measurements of high magnetic fields are performed  via the Zeeman splitting in nuclear magnetic resonance of hydrogen in water.  This technique based on radiofrequency/microwave frequency absorption, is applied mainly to continuous magnetic fields, with an uncertainty better than 1 ppm over a volume of a few mm$^3$. As well know from optical pumping, the detection of higher energy photons, {\it i.e.} as optical ones, greatly increases the measurement efficiency~\cite{Budker:2013}.  In presence of an applied magnetic field, atomic optical transitions experience a Zeeman frequency shift, and today laser frequency are measured with very high precision. The measure of a Zeeman shift is routinely applied to magnetic field in plasmas produced by an exploding wire~\cite{Garn:1966,Hori:1982,Gomez:2014,Banasek:2016}, where the high sample temperature limits the precision.\\
\indent We have developed an optical spectroscopy magnetic field probe based on the  Zeeman splitting in a rubidium atomic sample with a volume of 0.11 mm$^3~$\cite{George:2017}. The present investigation of atomic spectroscopy at high magnetic fields is based on that probe. Our experiment operates with pulsed magnetic fields having rise and fall times around 100 milliseconds. Even if  the detection is based on Doppler limited absorption spectroscopy, at the explored fields around 60 T the reached 20 ppm uncertainty allows us to perform high resolution optical spectroscopy. \\
\indent We report a precise study  of the Zeeman effect for the rubidium $5^2S_{1/2} \to 5^2P_{3/2}$ resonance line in magnetic field regimes not well explored.  Our results demonstrate that under a high resolution investigation the classification of the regimes as hyperfine or fine Paschen-Back ones~\cite{Kopfermann:1958}  represents a rough schematization for the atomic response. The data evidence that the hyperfine Paschen-Back regime is fully reached at magnetic fields larger than the standard comparison between electronic Zeeman energy and hyperfine structure splitting. On the other side, a theoretical description based on the fine Paschen-Back  approach is required to interpret data collected at magnetic fields lower than the standard comparison between electronic Zeeman energy and fine  structure splitting. \\
\indent As original feature, our measurement does not rely on the presence of an independent magnetometer, and the magnetic field value is derived directly from the measured rubidium optical absorption. The determination is based on the existence of an  optical Zeeman shift characterized by a field linear dependence, at all magnetic field values.  That measurement combined with the magnetic field temporal evolution detected  by a pick-up coil provides the absolute scale for the whole explored magnetic range. All atomic constants determining the rubidium absorption frequencies are well known from previous investigations, except for the Land\'e $g$-factor of the excited 5P state. Within the target of using the rubidium optical transitions for an atomic magnetometer, its precise value is required. We have derived the excited state Land\'e $g$-factor  by exploring the magnetic field dependence of different optical transitions. The ratio of the associated resonance fields, independent on the magnetic field absolute calibration, provides this  atomic constant with high precision. The accuracy of the rubidium based magnetometry is limited to 75 ppm by our $g$-factor uncertainty.\\
\indent Section II  describes the response to high magnetic fields, presenting eigenstates, eigenenergies and  optical transitions starting from the hyperfine Paschen-Back approximation.  The diamagnetic  contribution to the rubidium energy levels of interest is discussed.  We introduce the  optical transition line used for the magnetic field calibration. The Section is completed by a brief discussion on the atomic magnetic data required for our analyses. Section III  describes the experimental set-up, with the absorption detection based on the fluorescence detection in order to decrease the observation volume.   Section IV reports examples of recorded spectra, their analysis and the role played by the nuclear interaction. This Section includes also the analysis of the ratio of the high/low magnetic resonant fields for the $\sigma^+$ lines. The modification of the full fine structure multiplet is required for the data interpretation leading us to the Land\'e $g$-factor determination. A final Section concludes  our work.

\section{Rubidium in high magnetic field}
\subsection{Paramagnetism}
\label{Paramagnetism}
\begin{figure}
\centering
\includegraphics[width=7 cm]{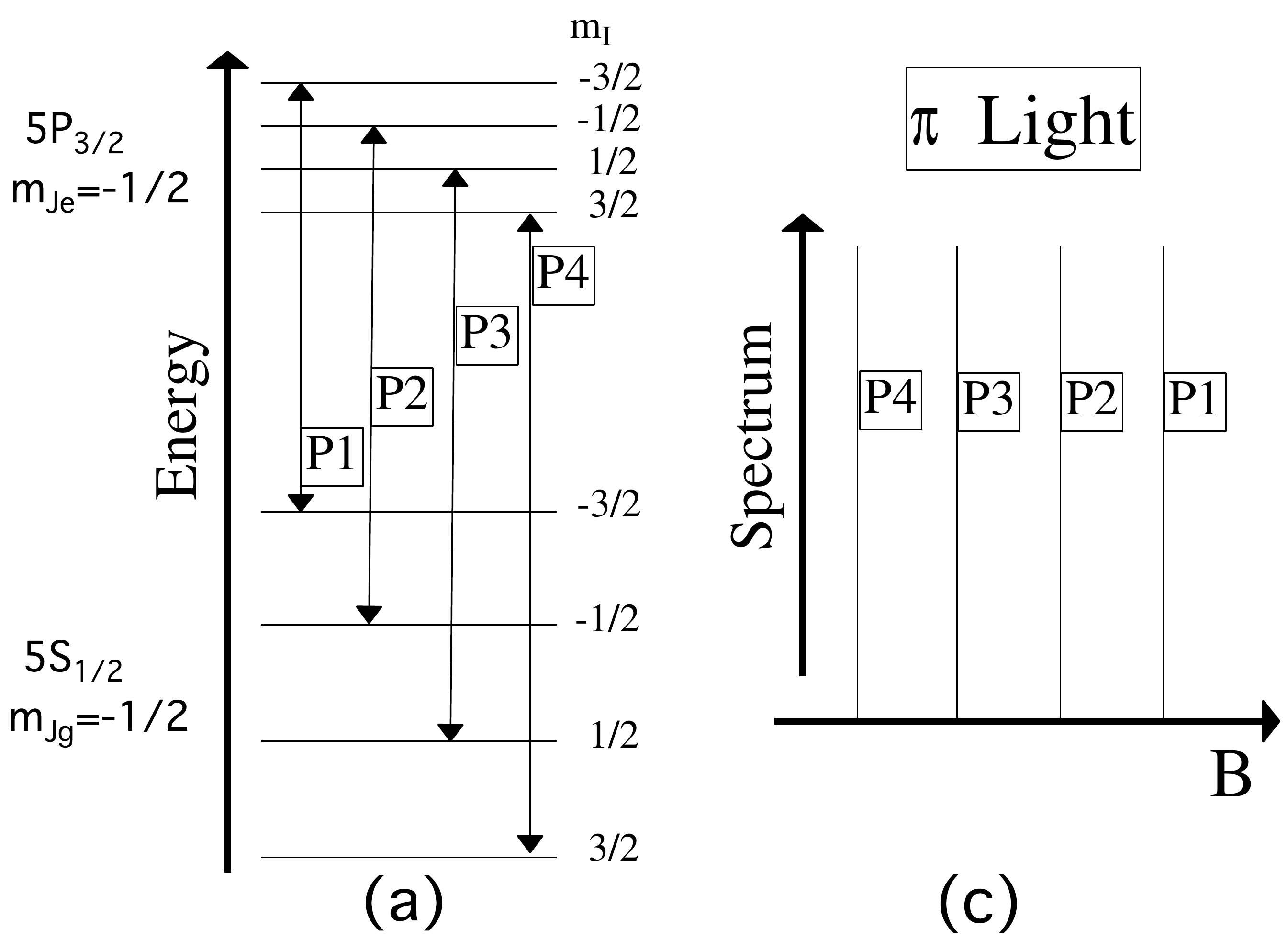}
\includegraphics[width=7 cm]{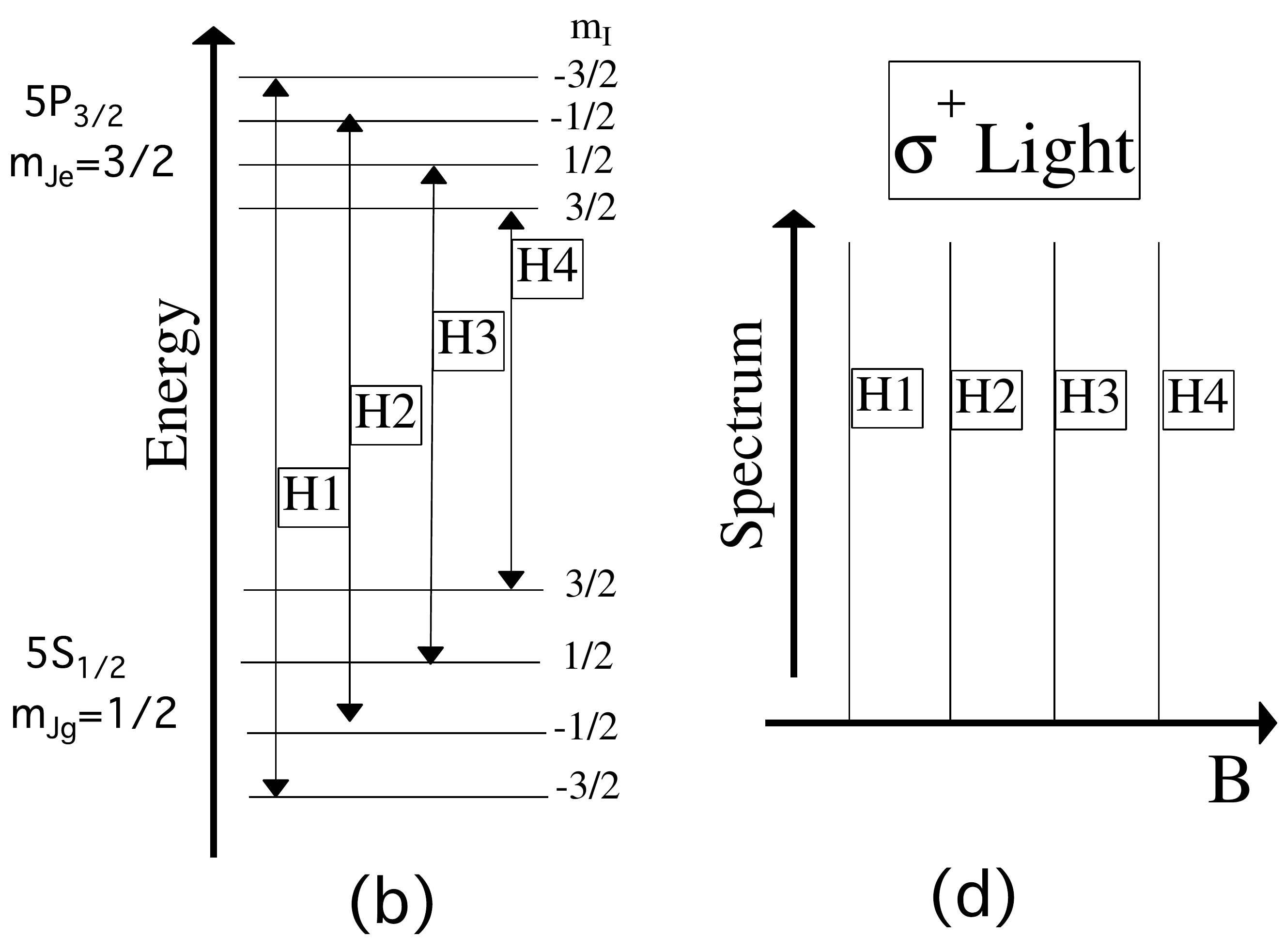}
\caption{In (a) and (b) energy levels at a given $B$ value. In (c) and (d) the spectrum produced by  transitions between the levels of (a) and (b), respectively, at a given laser frequency and scanning the $B$ magnetic field. The top and bottom plots correspond to the nuclear components of different electronic ground/excited states. (a) and (c)  correspond to the  $|J_g=1/2,m_{Jg}=-1/2\rangle \to |J_e=3/2,m_{Je}=-1/2\rangle$ $\pi$ transitions, denoted as $\pi$Pi with $(i=1,4)$.  (b) and (d)  correspond to the  $|J_g=1/2,m_{Jg}=1/2\rangle \to |J_e=3/2,m_{Je}=3/2\rangle$ $\sigma^+$ transitions, denoted as $\sigma^+$Hi.  Notice the energy order of the nuclear levels for different electronic states, producing a different order of the $\pi$Pi and  $\sigma^+$Hi  lines while  scanning $B$.}
\label{energiesspectra}
\end{figure}

For an alkali atom, the magnetic interaction of the valence electron is described through several quantum numbers,  for the  nucleus the $(I,m_I)$ spin components, for the electron the $(L,m_L)$ orbital angular momentum, the $(S,m_S)$ spin, the  vector composition of  $\mathbf{L}$ and $\mathbf{S}$ with $(J,m_J)$ components, and finally the vector composition of $\mathbf{J}$  and $\mathbf{I}$ with components $(F,m_F)$. As in textbooks~\cite{Kopfermann:1958}, at very low magnetic fields, where the electron-nucleus hyperfine interaction is larger than the electronic Zeeman interaction,   $(F,m_F)$  are  the correct quantum numbers. Increasing the magnetic field the hyperfine Paschen-Back regime is reached when the role between hyperfine interaction and electronic Zeeman energy is reversed~\cite{Kopfermann:1958}. There  $(J,m_J)$ are the good quantum numbers. Because for rubidium the  $5^2S_{1/2}$ ground state hyperfine interaction is around fifty times larger than the $5^2P_{3/2}$  excited state one, the  hyperfine Paschen-Back  regime should be reached when the electron Zeeman energy is roughly equal to the the ground state interaction, around 0.5 T for $^{87}$Rb. While for the ground state $(J,m_J)$ remain  good quantum numbers for all higher magnetic fields, for the excited state the fine structure splitting between $5^2P_{1/2}$ and $5^2P_{3/2}$ states must be compared to the electronic Zeeman splitting. For rubidium the fine Paschen-Back regime is reached at fields around 500 T.  There  $(L=1,m_L)$ and  $(S=1/2,m_S)$ are the quantum numbers, combined with the nuclear spin.  This regime was never explored in rubidium, while owing to the smaller fine structure splitting it was explored for sodium in experiments with exploding wires~\cite{Garn:1966,Hori:1982,Gomez:2014,Banasek:2016}.\\
\indent  At high fields the diamagnetism contribution to the magnetic interaction should be included in the analysis, as in Sec IIC. However our experiment demonstrates this contribution negligible for fields up to 60 T.  For completeness refs.~\cite{LipsonLarson:1986,Fortson:1987} explored for the rubidium ground state a Zeeman energy term induced by  the magnetic dipole hyperfine interaction, through a coupling of higher electronic levels  into the ground state, term quadratic in $m_I$ and equivalent to a modification of the nuclear $g$-factor. This shift is only few Hz at our largest explored field.\\
\indent Following the above classification, the  1-60 T explored magnetic field range corresponds to the  hyperfine Paschen-Back  regime.  Thus the rubidium ground state is specified by   $(J_g=1/2,m_{Jg})$, combined with $(I,m_I)$. The two stable isotopes, $^{85}$Rb and $^{87}$Rb, have  nuclear spin, $I=5/2$ and $I=3/2$, respectively, characterized by the nuclear Land\'e $g$-factor $g_I$, assumed  negative as in~\cite{Arimondo:1977}. For the alkali ground state, the eigenenergies are given by the Breit-Rabi formula~\cite{Kopfermann:1958,Foot:2012}, including the electronic and nuclear Zeeman energies and  the  $A_g$ dipolar hyperfine coupling. \\
\indent For the excited state, no analytical formula exist for the eigenenergies,  to be derived  by diagonalizing numerically the Hamiltonian.   For the  hyperfine Paschen-Back  regime the Hamiltonian contains the magnetic interactions and the hyperfine dipolar ($A_e$)  and quadrupolar  ($B_e$)  contributions. For the fine Paschen-Back regime,  the Hamiltonian includes the above contributions and the fine structure splitting of the excited 5$^2P_{1/2,3/2}$ states.\\
\indent For an applied $B$ field and in the hyperfine Paschen-Back regime,  the energies of the ground and excited states, $E_g$ and $E_e$, respectively, expressed in frequency units are given by
\begin{eqnarray}
&E_g(J_g=1/2,m_{Jg};I,m_{Ig})=\mu_B\left(g_{5S}m_{Jg}+g_Im_{Ig}\right)B\nonumber\\
&+A_gm_{Jg}m_{Ig}, \nonumber \\
&E_e(J_e=3/2,m_{Je};I,m_{Ie})=\mu_B\left(g_{5P_{3/2}}m_{Je}+g_Im_{Ie}\right)B\nonumber \\
&+A_em_{Je}m_{Ie}+B_e\frac{6(m_{Je}m_{Ie})^2+3m_{Je}m_{Ie}-2I(I+1)J_e(J_e+1)}{4I(2I-1)J_e(2J_e-1)}.
\label{PaschenBackregime}
\end{eqnarray}
where $\mu_B$ is the Bohr magneton in MHz/T, $g_{5S}$ and $g_{5P_{3/2}}$ are the electronic  Land\'e $g$-factors. For the investigated magnetic field range the nuclear Zeeman contribution is larger than the hyperfine coupling for the excited state, and comparable for the ground state. That determines different dependences of the     state energy on the nuclear quantum number, as shown Figs.~\ref{energiesspectra} (a) and (b).  \\

\subsection{Absorption lines}
\indent  The energy levels and the optical transitions are here discussed for the $^{87}$Rb isotope with $I=3/2$. For a given magnetic field B and a given laser frequency $\nu_L$, the absorption spectrum is composed by lines at frequencies
\begin{equation}
\nu_L=\nu_0+[E_e(\frac{3}{2},m_{Je};\frac{1}{2},m_{Ie})-E_g(\frac{1}{2},m_{Jg};\frac{3}{2},m_{Ig})],
\label{absorption}
\end{equation}
where $\nu_0$ is the  rubidium absorption center of gravity at $B=0$,  Within the strong magnetic field regimes the light induces mainly transitions with $\Delta m_I=0$ selection rules. Because of the small diamagnetic contributions, as in Secs.~\ref{diamagnetism} and~\ref{spectra}, the position of  absorption lines is dominated by the paramagnetic contributions of Eqs.~\eqref{PaschenBackregime}.  \\
\indent  Our experimental approach is based on imposing an offset $\Delta \nu_L=\nu_L-\nu_0$  and pulsing the magnetic field from zero to a preset maximum value. At specific times the atoms reach a resonance with the laser by the Zeeman effect.  The  fluorescence emission monitors the atomic absorption. For our $\Delta \nu_L$ positive values, the spectra observed by scanning the magnetic field contain three sets of lines, denoted $\pi$Pi, $\sigma^+$Hi, and $\sigma^+$Li,  starting from a high resonant field to a low one, with $(i=1,4)$  corresponding to the $(m_I=\pm3/2,\pm1/2)$ components. \\
\indent The $\pi$ polarized Pi lines, produced by the  $|J_g=1/2,m_{Jg}=-1/2\rangle \to |J_e=3/2,m_{Je}=-1/2\rangle$ transitions, experience the smallest Zeeman shift. Fig.~\ref{energiesspectra}(a)  shows the energy levels corresponding to these  transitions at given $B$.  Fig.~\ref{energiesspectra}(c) schematizes their  nuclear structure as observed at fixed $\Delta \nu_L$  and scanning $B$. \\
 \indent  Fig.~\ref{energiesspectra}(b)  shows the energy levels corresponding to the  $|J_g=1/2,m_{Jg}=1/2\rangle \to |J_e=3/2,m_{Je}=3/2\rangle$ $\sigma^+$Hi polarized transitions at fixed $B$.  The ground state nuclear structure is dominated by the hyperfine interaction even at the highest explored magnetic field, and produces an opposite ranking of the $m_I$ levels in Fig.~\ref{energiesspectra}(a) and (b) because of the different $m_{Jg}$ sign.  The excited state nuclear structure, dominated by the nuclear Zeeman effect, is the same for the two cases.  As a consequence scanning the $B$ field, the order of the lines is opposite for  the $\pi$Pi and $\sigma^+$Hi cases, as in Fig.~\ref{energiesspectra}(c) and (d). The centers of gravity of the absorption lines eliminating the hyperfine structure contribution play a key role in our data analysis. For both   hyperfine and fine  Paschen-Back regimes the center of gravity for the $\sigma^+$Hi transitions  is
\begin{equation}
B^H_{\rm Center}=\frac{\Delta \nu_L}{\mu_B(3g_{5P_{3/2}}-g_{5S})/2}.
\label{BHcenter}
\end{equation}
 \indent Similar $\sigma^+$ transitions denoted $\sigma^+$Li  are the  $|J_g=1/2,m_{Jg}=-1/2\rangle \to |J_e=3/2,m_{Je}=1/2\rangle$  ones. These transitions appear at a magnetic field lower than the previous ones, because they experience a larger Zeeman shift.  The order of the $\sigma^+$Li lines observed while scanning the magnetic field is reversed in respect to that of the $\sigma^+$Hi lines and similar to the $\pi$Pi ones.  For the hyperfine Paschen-Back  regime only, the center of gravity of the $\sigma^+$Li transitions is
 \begin{equation}
B^L_{\rm Center}=\frac{\Delta \nu_L}{\mu_B(g_{5P_{3/2}}+g_{5S})/2}.
\label{BLcenter}
\end{equation}
 Within the  hyperfine and fine Paschen-Back regimes, the  $\pi$Pi, $\sigma^+$Hi, and $\sigma^+$Li nuclear transitions are not  equally spaced, because of the  small excited state hyperfine quadrupole coupling. \\
\begin{figure}[b!]
\centering
\includegraphics[width=8 cm]{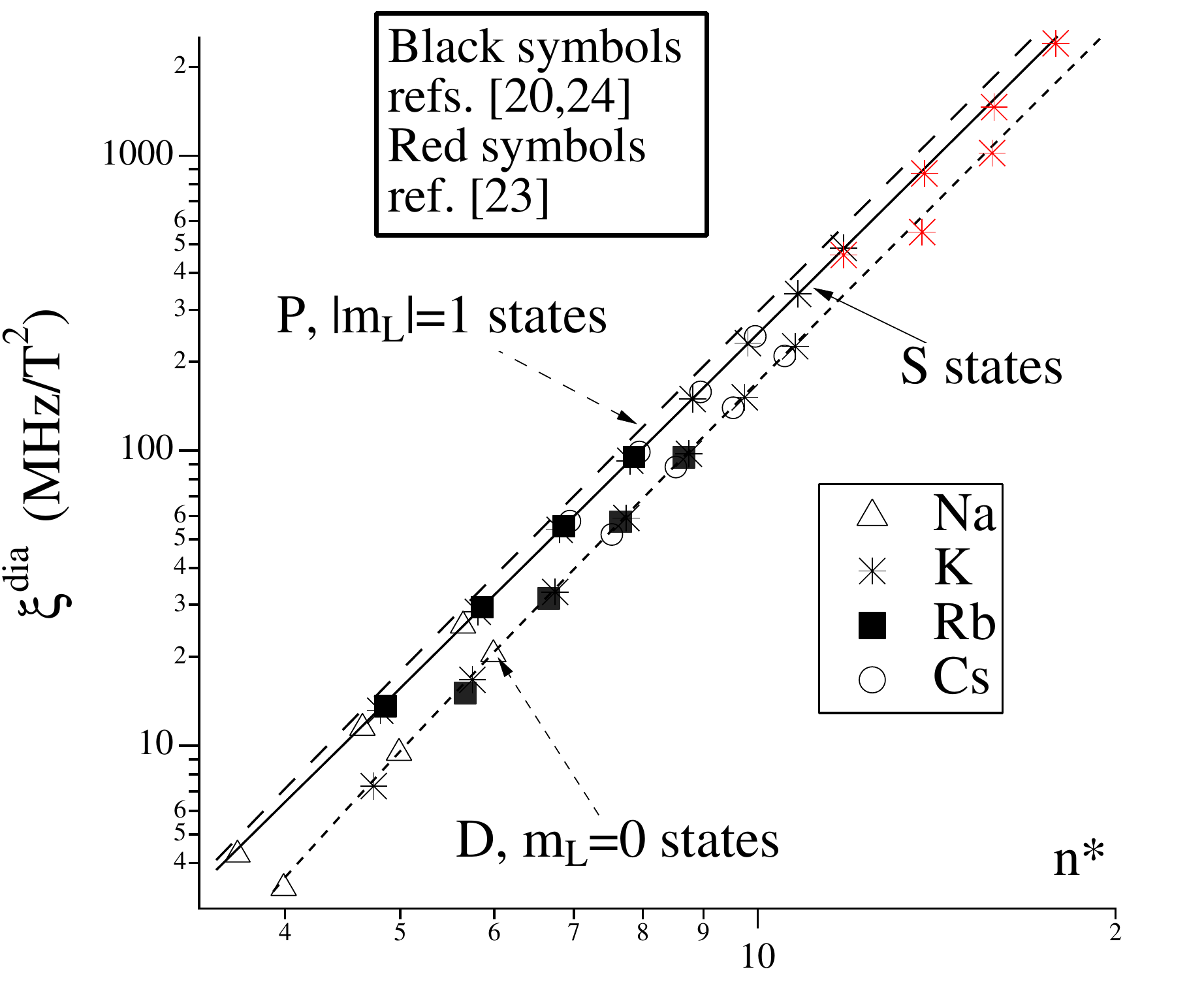}
\caption{Experimental results and theoretical predictions of $\xi^{dia}$ vs the $n^*$ effective quantum number  for $S$, $(P, m_L=\pm 1)$  and $(D, m_L=0)$ low energies states of alkalis.  Eq.~\eqref{DiaSuscept} predictions for those states correspond to the continuous, dashed and dotted lines, respectively. Triangles, stars, filled squares and open circles refer to Na, K,  Rb and Cs results, respectively, measured in the references reported within the top box. For our $5S$ and $5P$ states the $n^*$ values are 1.81 and  2.30, respectively.}
\label{FigDiamagnetism}
\end{figure}
\subsection{Diamagnetism}
\label{diamagnetism}
Diamagnetic corrections are necessary for accurate measurements at  high magnetic fields. If the magnetic perturbation is smaller than the energy separation between states with different $L$ quantum numbers,  the $E^{dia}$ diamagnetic energy for a single valence electron may be written
\begin{equation}
E^{dia}=\xi^{dia}B^2 \\
\label{DiaEnergy}
\end{equation}
on the basis of the susceptibility $\xi^{dia}$. Within an hydrogen-like description, for an electron with quantum numbers $n,L,m_L$, and an effective quantum number $n^*=n-\delta$   determined by the quantum defect $\delta$, $\xi^{dia}$ derived by~\cite{SchiffSnyder:1939,Garstang:1977} was  rewritten by~\cite{Otto:2002} as
\begin{equation}
\xi^{dia}=\frac{5e^2a_0^2}{8\mu_e}\left[1+\frac{1-3l(l+1)}{5(n^*)^2}\right]
\frac{l(l+1)+m_l^2-1}{(2l-1)(2l+3)}(n^*)^4,
\label{DiaSuscept}
\end{equation}
with $a_0$ the Bohr radius, $e$ the electron charge and $\mu_e$ the electron reduced mass.\\
\indent Fig.~\ref{FigDiamagnetism}  reports experimental determinations and  theoretical predictions for $\xi^{dia}$ vs $n^*$ in alkalis for low $n$ quantum numbers. Early measurements  were performed on P states in Na and K at 2.7 T~\cite{JenkinsSegre:1939} and in K, Rb, Cs  at 2.3 T~\cite{HartingKlinkenberg:1949}.  The diamagnetic contribution was measured using  two-photon spectroscopy excited S and D states originally by~\citep{HarperLevenson:1977}, and later more systematically for all alkalis by~\cite{Huettner:1996,Otto:2002}.    The measurements by~\cite{Otto:2002} for $S$ and $D$ states on different alkalis span $n$ quantum numbers as low as three. The simple hydrogen atom description of Eq.~(\eqref{DiaSuscept}) with quantum defects derived from~\cite{Gallagher:1994,LiGallagher:2003} provides a good fit to the susceptibilities of the figure. The $P$ states measurements by~\cite{JenkinsSegre:1939,HartingKlinkenberg:1949} on several alkalis focused  on quantum numbers between 12 and 30.  Those low precision data  are also  fitted by  that equation, except that for Rydberg states with $n>20$ the inter-$L$ perturbation interactions introduce deviations from the predicted values.\\
\indent Eq.~\eqref{DiaSuscept} predicts the $\xi^{dia}$ diamagnetic values of 0.29 MHz/T$^2$ and 0.68 MHz/T$^2$, respectively,  for our $5S$ and $5P$ states. That  leads to a predicted $\Delta \nu^{dia}=0.39$ MHz/T$^2$ diamagnetic shift for the $\sigma^+$H4 line, compared to its $\approx14$ GHz/T paramagnetic shift.   All these  values should be considered only as an order of magnitude,  because of the limited validity of  the hydrogen-like  description for our states with the low $n^*$  values reported in the figure caption. At 58 T the predicted diamagnetic shift is around 1.3 GHz, corresponding to about three times the Doppler linewidth of the absorption lines, whence measurable. Our $\xi^{dia}$ result is  presented at the end of Sec.~\ref{spectra}. \\

\subsection{Data analysis}
\label{dataanalysis}
\indent  Our  $^{87}$Rb spectral analysis is based on the atomic constants reported in~\cite{Steck87:2001}, except for the Land\'e $g$-factors.  For $^{87}$Rb the ground state Land\'e g-factor was precisely measured in~\cite{Tiedeman:1977} with respect to the  $g_e$ free electron $g$-factor.  Making use of the $g_e$ value given in~\cite{CODATA:2016}  we obtain $g_{5S} = 2.002331070(26).$ A larger indetermination is associated to the Land\'e $g_{5P_{3/2}}$-factor.  The data of~\cite{Arimondo:1977} point out that for all the alkali atoms the $g$-factor of the first excited $P_{3/2}$ state is  $\approx 1.33411$, as predicted by the Russel-Saunders coupling between the orbital magnetic moment, with $g_L=1$, and the spin magnetic moment, using $g_e$ or  $g_{5S}$.  For the $^{87}$Rb $5P_{3/2}$ state, ref.~\cite{Arimondo:1977} reported 1.3362(13) as a weighted average of all the measurements available at that time and still today. That value is largely determined by fitting the level crossing measurements by Belin and Svanberg~\cite{BelinSvanberg:1971}, who derived simultaneously $g_{5P_{3/2}}$ and the dipolar and quadrupolar hyperfine constants. We have reanalysed those level-crossing measurements  by fixing the hyperfine constants  to the very precise values of ref.~\cite{YeHall:1996} reported in~\cite{Steck87:2001} and using the $^{87}$Rb nuclear magnetic moment of ref.~\cite{Arimondo:1977}.  A new $g_{5P_{3/2}}= 1.3341(2)$ value is obtained, in agreement with the above Russel-Saunders prediction, to be used as the starting point of our analysis.  Following ref.~\cite{Labzowsky:1999,Goidenko:2002,Indelicato:2007,GosselFlambaum:2013}, QED and relativistic corrections  are at the level of 10$^{-4}$-10$^{-5}$.

\subsection{Rb atom as magnetometer}
\indent Our magnetic field determination is based on the rubidium spectrum itself. It relies on the existence of two eigenstates,  ground and excited, denoted as extreme, whose energy dependence on the magnetic field is exactly linear, excluding the diamagnetic contribution. These eigenstates correspond to the highest values of all the atomic quantum numbers. The  $5^2S_{1/2}$ $|J_g=1/2,m_{gJ}=1/2;I,m_{gI}=I\rangle$ ground state  has the following energy:
\begin{equation}
E_g^+ =\mu_B\left(\frac{g_{5S}}{2}+g_II\right)B+\frac{1}{2}A_gI.
\label{EstrGr}
\end{equation}
The excited eigenstate with the highest energy, i.e., the $5P_{3/2}$ $|J_e=3/2,m_{eJ}=3/2;I,m_{eI}=I\rangle$ state has the following energy whichever magnetic field value:
\begin{equation}
E_e^+ =\mu_B\left(\frac{3g_{5P_{3/2}}}{2} + g_II\right)B +\frac{3}{2}A_eI+\frac{1}{4}B_e.
\label{EstrExc}
\end{equation}
These formula for the extreme states, even if derived from Eqs.~\eqref{PaschenBackregime} valid in the hyperfine Paschen-Back  regime, apply to all regimes, even for the  fine  Paschen-Back one.\\
\indent Combining together Eqs.~\eqref{absorption},~\eqref{EstrGr} and ~\eqref{EstrExc}, the Zeeman frequency shift of the $\sigma^+$H4 optical transition linking the Rb linear dependent states is given by
\begin{equation}
\Delta \nu_{L}(\sigma^+H4)=\mu_BB\frac{3g_{5P_{3/2}}-g_{5S}}{2}+ \frac{I}{2}(3A_e-A_g)+\frac{1}{4}B_e.
\label{nu}
\end{equation}\\
The inversion of this equation allows to derive the $B$ value from the laser frequency exciting the Rb atoms.\\
\indent For our Doppler limited spectroscopy with the Gaussian absorption center determined at one twentieth of its linewidth, the magnetic field precision  is~$\approx 0.002$ T.  The above determination leads to the 20 ppm precision at high fields. The Rb magnetometry accuracy is determined by the $g_{5P3/2}$ uncertainty, 750 ppm for the value reported in~\cite{Arimondo:1977} and 75 ppm for the value derived in Sec. IV.

\begin{figure}
\centering
\includegraphics[width=8 cm]{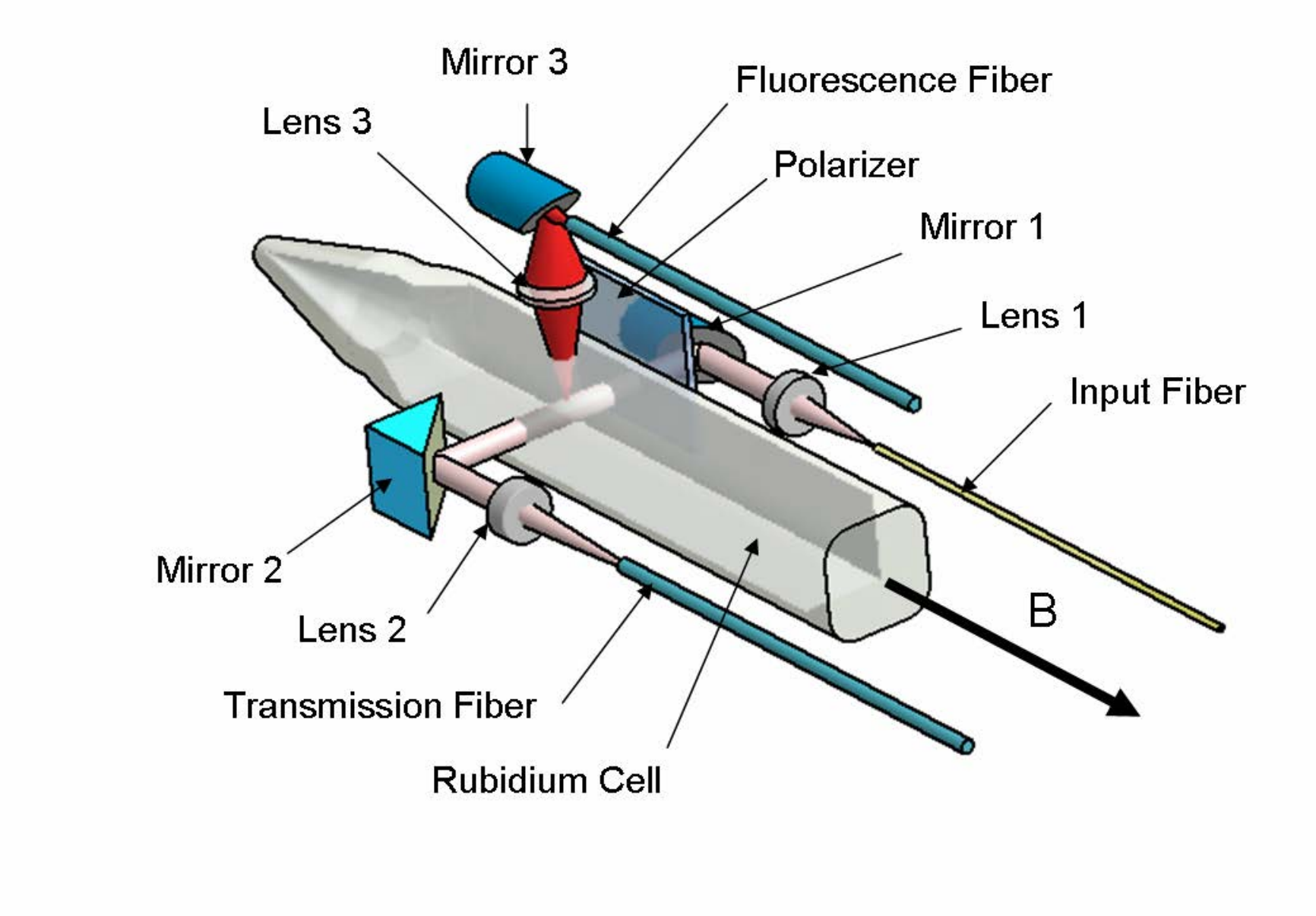}
\includegraphics[width=9 cm]{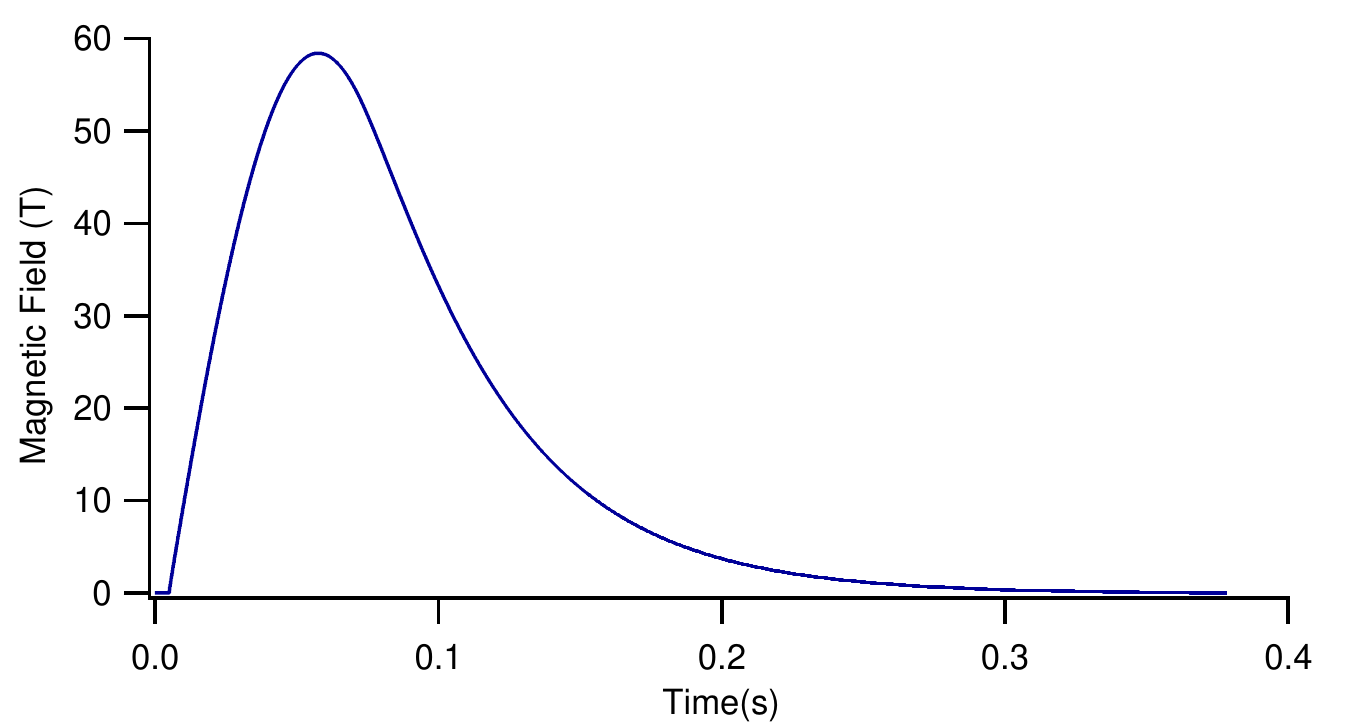}
\caption{On the top, optical scheme of the Rb cell with a single mode fiber for the laser input  and multimode fibers for the transmission light and the fluorescence light output. Matching lenses and a  45$^{\circ}$ linear polarizer controlling the light reaching the Rb atoms are shown. The cell is oriented parallel to the $B$ field direction and the interrogation light propagates orthogonally to the field. On the bottom, magnetic field temporal dependence for a typical pulse.}
\label{sensor}
\end{figure}

\section{Probe and magnet}
\label{sec:setup}
\indent The experimental set-up is composed by the rubidium probe located at the center of a solenoid magnet, immersed into liquid nitrogen. The Rb probe is composed by a quartz cell located at the end of a long pipe placing the cell within the magnet and hosting all the electrical and optical connections. The cell with   $3\times3\times30$ mm$^3$ internal dimensions is filled with natural rubidium. To maintain the cell at room temperature the atomic probe is placed within an evacuated  double-walled stainless steel cryostat inserted into the magnet bore. As in the top of Fig.~\ref{sensor}, a single mode optical fiber provides the input beam, while the transmitted light and the atomic fluorescence emission reach the outside detectors through multimode fibres.  Before entering into the cell the light, generated by a DLX100 Toptica laser, is polarized at 45$^{\circ}$ with respect to the magnetic field direction in order to induce both $\pi$ and $\sigma$ transitions. The Faraday rotation, experienced by  the light propagating through the input single mode fiber and parallely to the magnetic field direction, modifies the total intensity, not the polarization on the atoms. The reported data for a laser intensity fifteen times the saturation  intensity are produced by a few ten thousand atoms. Previous tests, performed before assembling the cell within the magnet, demonstrated that the minimum number of detectable atoms is around 100. The fluorescence observation instead of the transmission allows to reduce the influence of the magnetic field inhomogeneity. In fact  from the fiber diameter and the collection lens parameters, we evaluate that the fluorescence light is produced from a  $0.4\times0.4\times0.7$ mm$^3$ volume, smaller than the cell volume probed by transmission. \\
\indent The 60 T pulsed magnetic field coil, a standard one at the High-Field National Laboratory (LNCMI) in Toulouse, has a 28~mm free bore diameter and is  is immersed into the liquid nitrogen in order to facilitate the heat dissipation~\cite{Debray:2013}.  The magnetic field homogeneity on the probed atomic volume is estimated better than 10~ppm. The risetime and decaytime of the field temporal evolutions are around 55~ms and  100~ms, respectively, as shown in the bottom of Fig.~\ref{sensor}.  The field temporal evolution is monitored by a pick-up coil located at 7 mm from the atoms.  Its frequency response bandwidth is larger than 500 kHz. The pick-up signal is calibrated in a separate carefully designed solenoid.  The integrated pick-up signal reproduces the time profile of the magnetic pulse. That signal, corrected for the distance from the probe center position, provides a reference measurement $B_{PU}$ of the magnetic field experienced by the atoms. The field calibration is based on the $B_{H4}$ theoretical prediction for the $\sigma^+$H4 resonance derived from Eq.~\eqref{nu} at a given laser frequency. From the analysis of $\approx$70 spectra, a  linear dependence between $B_{PU}$ and $B_{H4}$ was verified, with slope $ 0.9899(2)$ using the Russell-Saunders $g$-factor reported above.  For a more detailed set-up description see~\cite{George:2017}.

\begin{figure}[b!]
\centering
\includegraphics[width=9 cm]{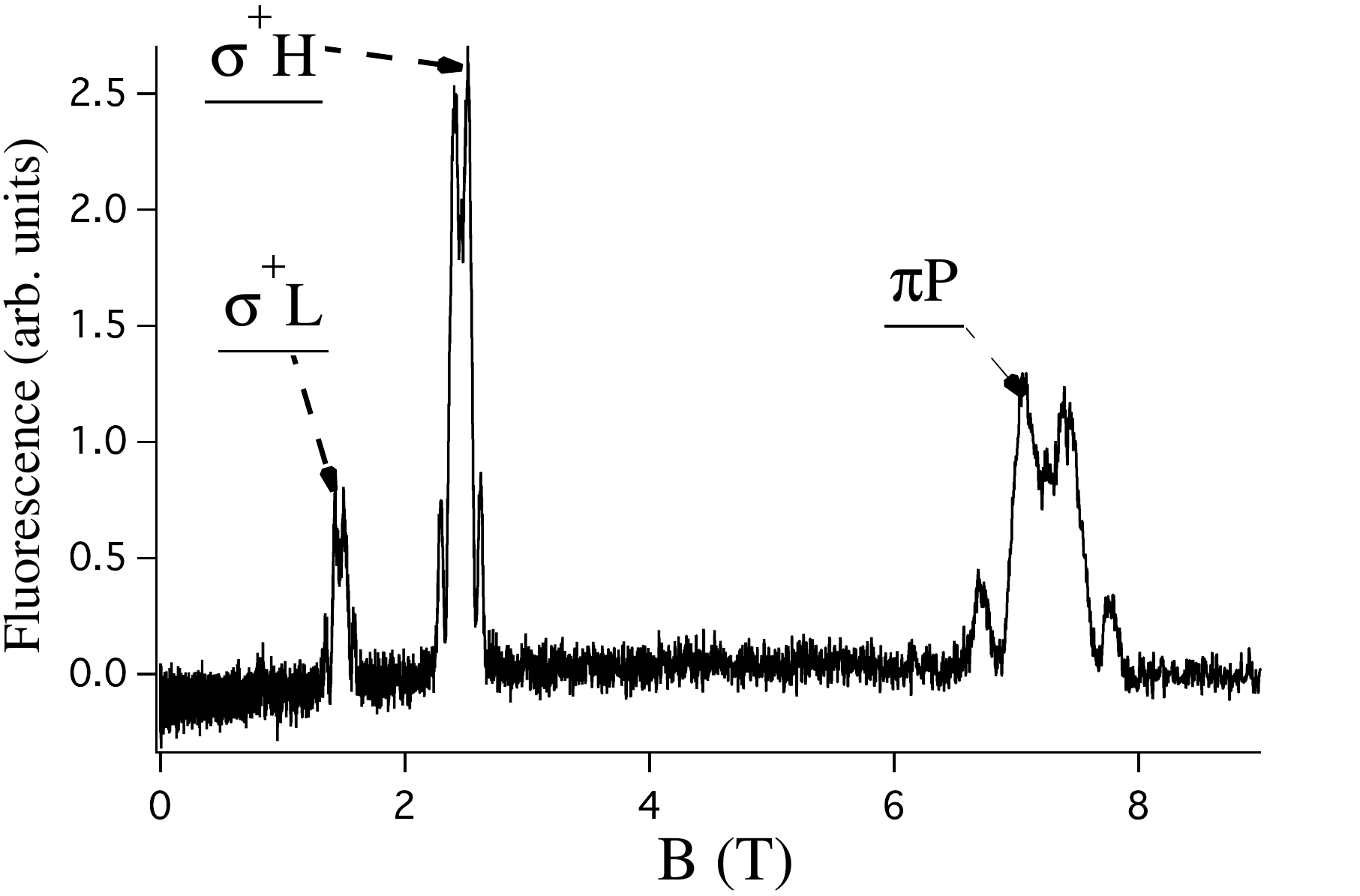}
\includegraphics[width=9 cm]{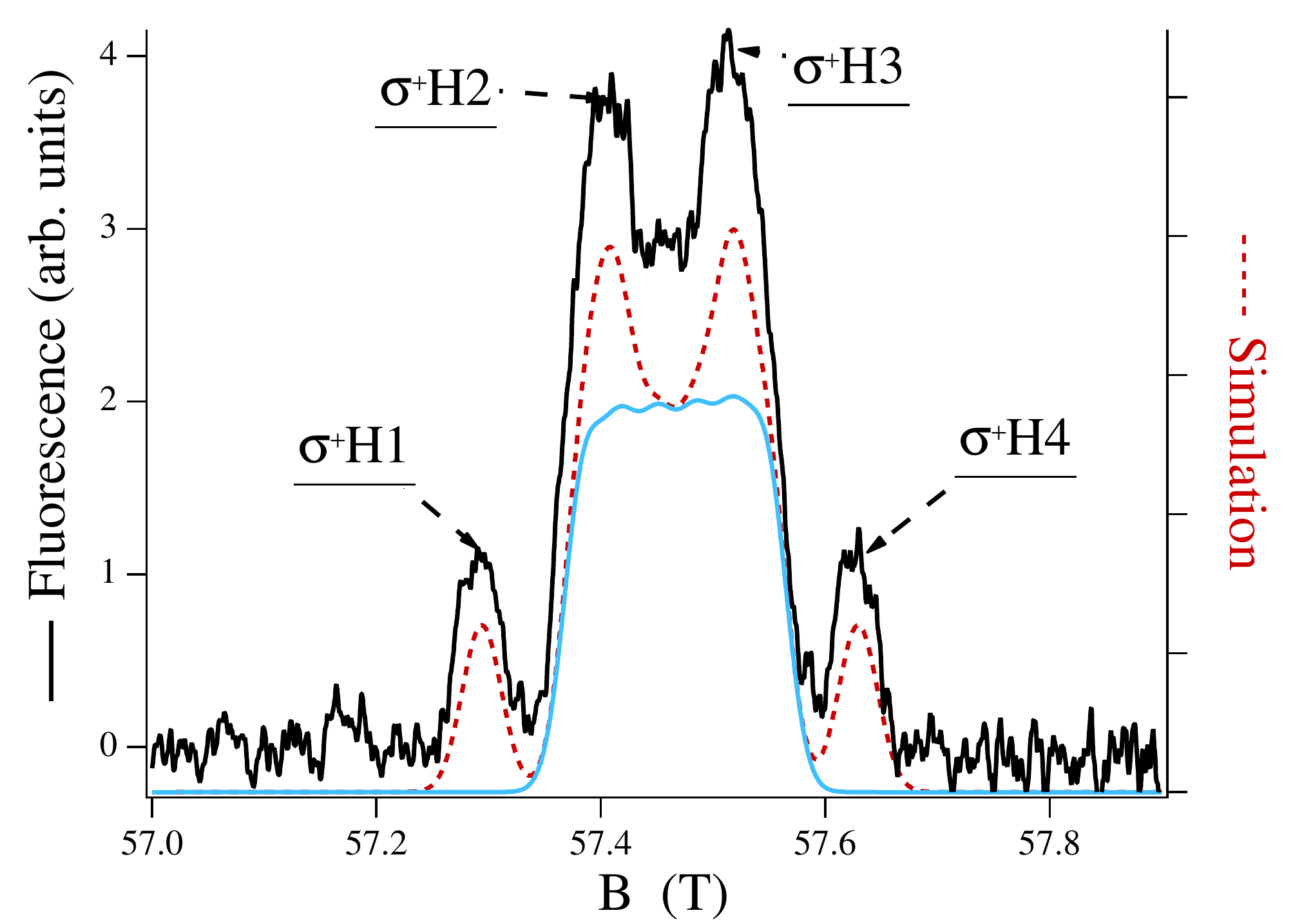}
\caption{Rb fluorescence spectra observed by scanning the magnetic field, on the top for  $\Delta \nu_L=34.692(5)$  and on the bottom for $\Delta \nu_L=812.331(5)$, all in GHz, with uncertainty given by the accuracy of the wavelength meter reading. For both spectra the sinusoidal periodic fluorescence variation produced by the fiber Faraday effect was subtracted. The bottom plot, on an expanded scale, shows the individual $\sigma^+$Hi lines for a very high magnetic field. The quantum number assignment is in Fig.~\ref{energiesspectra}. Each fluorescent structure is composed by four resolved lines associated to the $^{87}$Rb nuclear spin states and by a central unresolved broad structure associated to the $^{85}$Rb nuclear components.  The dotted red line reports a simulation including the Gaussian Doppler broadening for each absorption line, while the blue continuous line reports only the $^{85}$Rb contribution, with an offset  for both of them, for presentation clarity.  The simulation only free parameter is the overall scale. }
\label{fluorescencespectra}
\end{figure}

\section{Experimental results}
\subsection{Spectra}
\label{spectra}
Examples of the observed fluorescence spectra are reported in Fig.~\ref{fluorescencespectra} for two different values of the $\Delta \nu_L$ while the magnetic field is scanned during the pulse  decaytime. Similar spectra are obtained while the magnetic field is scanned up. The top spectrum, obtained for a  low  $\Delta \nu_L$, is characterized by the presence of all the $\sigma^+$Li, $\sigma^+$Hi and $\pi$Pi  lines  in sequence at increasing $B$  values. The lines at higher $B$ values  experience a smaller Zeeman shift.  The intensities are proportional to the theoretically predicted line strengths. In the bottom spectrum, obtained for a large $\Delta \nu_L$, the $\sigma^+$Hi  lines appears for a magnetic field  close to the coil maximum operational current. The horizontal scales are obtained combining the information provided by the $B_{H4}$ value and the temporal magnetic field dependence measured by the pick-up coil.\\
\indent Each fluorescent set  includes the  $^{85}$Rb and $^{87}$Rb contributions. The four peak structure observed for each set  corresponds to the nuclear structure of the $^{87}$Rb $I=3/2$ spin. Because for $^{85}$Rb the $A_g$  hyperfine coupling is two times smaller than for   $^{87}$Rb, and because of its higher spin value $I=5/2$, the nuclear structure cannot be resolved by the Doppler limited spectroscopy.  We have performed simulations of the spectra including the Doppler Gaussian broadening, an example represented by the red  dotted line of  Fig.~\ref{fluorescencespectra} bottom. The simulation reproduces the four $^{87}$Rb $\sigma^+$Hi lines  having the same intensity and the central broadening due to the unresolved $^{85}$Rb lines. On the bottom spectrum,  the asymmetry between the two sides of the absorption structure, clearly visible on the $^{85}$Rb simulation, is produced by the unequal spacing among the nuclear levels, around 0.002 T in the resonant magnetic field.\\
\indent Analysing spectra obtained for  different $\Delta \nu_L$ values, we derive the linear relation between $B_{PU}$ and $B_{H4}$ reported within Sec.~\ref{sec:setup}. In order to test the presence of a quadratic diamagnetic  nonlinearity, we repeat the previous analysis of $B_{PU}$ vs $B_{H4}$ by including into the fit function a  quadratic term. The fit quality is not improved and the  derived  $ \xi^{dia}$  value is smaller than the above theoretical prediction by a factor ten and compatible with zero owing to a large error bar.\\

\begin{figure}[t!]
\centering
\includegraphics[width=9 cm]{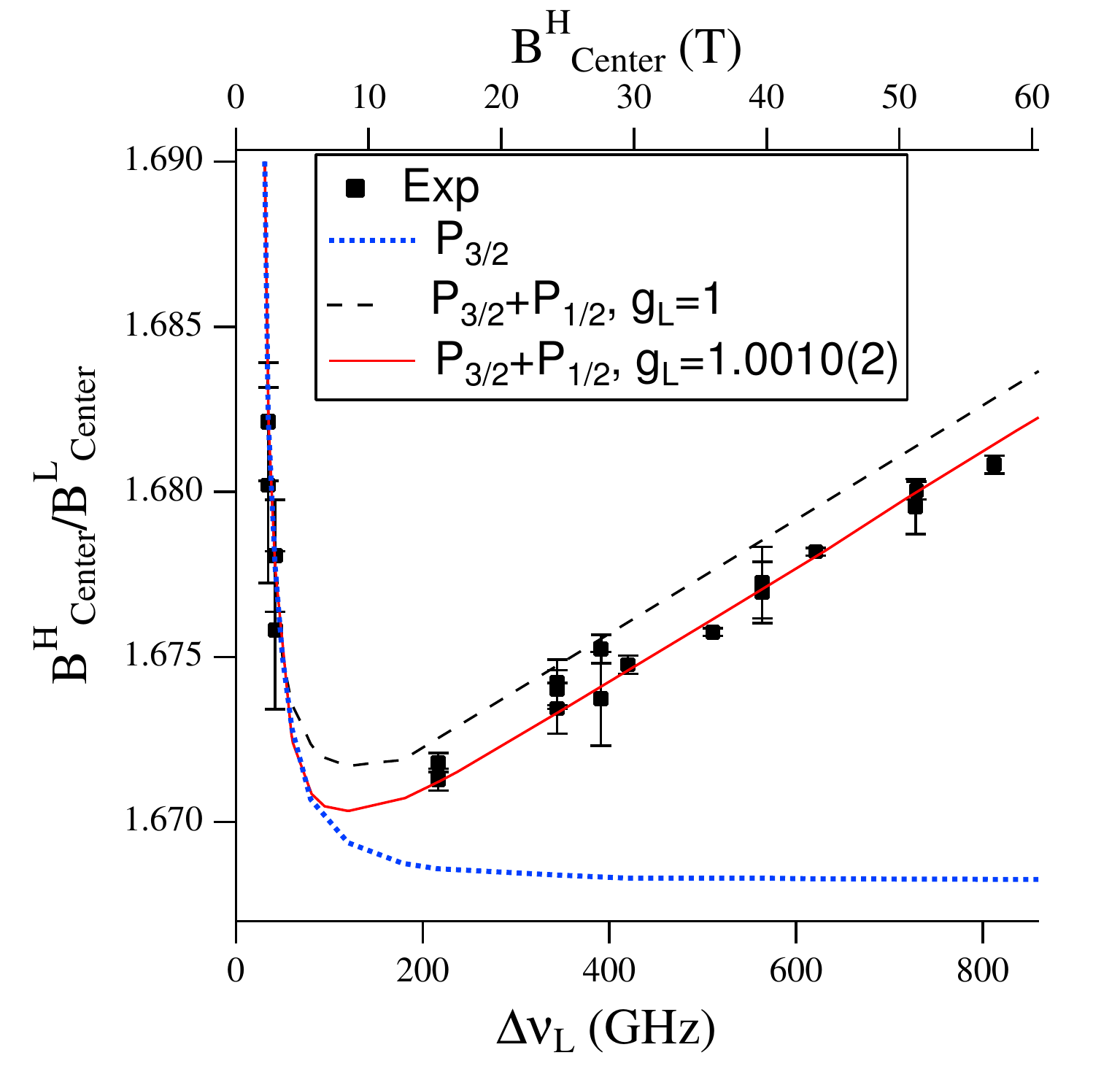}
\caption{Ratio between $BH_{\rm Center}$ and $BL_{\rm Center}$ magnetic fields versus the $\Delta \nu_{L}$ laser detuning (bottom axis) and $BH_{\rm Center}$ value (top axis). Squares for the measured values, including error bars, in several cases smaller than the square. Lines report theoretical predictions. The  dotted blue line is based on the 5$^2$P$_{3/2}$ eigenergies, at high magnetic fields leading to the hyperfine Paschen-Back  regime with a constant value of the ratio. The black dashed line and the red continuous line are based on  the magnetic eigenenergies of the whole fine structure   5$^2$P$_{3/2}$ and  5$^2$P$_{1/2}$ manifold, describing at high fields the fine Pachen-Back  regime. Different $g_L$-values, shown in the inset,  with $g_S=g_{5S}$, are used for the two cases. The red line results are obtained also for other $[g_L$, $g_S]$ combinations discussed in the text.   }
\label{gjratio}
\end{figure}

\subsection{Ratio between $B^H_{\rm Center}$ and $B^L_{\rm Center}$ }
This Subsection targets the $g_{5P_{3/2}}$ value that among the Rb data  has a low accuracy. Eqs.~\eqref{BHcenter} and ~\eqref{BLcenter} show the different dependence of $B^H_{\rm Center}$ and $B^L_{\rm Center}$ on $g_{5P_{3/2}}$ because of the different excited state $m_{Je}$ quantum numbers. Those equations don't have the same regime of validity: the $B^H_{\rm Center}$ expression is valid at all fields; the $B^L_{\rm Center}$ is based on the hyperfine Paschen-Back approximation. Within this approximation, at high magnetic fields, for a given $\Delta \nu_L$ laser detuning the $B^H_{\rm Center}/B^L_{\rm Center}$ ratio is a constant. This prediction is shown by the dotted blue line in Fig.~\ref{gjratio} where also  the experimental results for  the ratio are plotted as function of  $\Delta \nu_L$  on the bottom axis, or the corresponding $B^H_{\rm Center}$  center  on the top axis. Notice the high precision reached by the measurements at very high fields, where the Doppler linewidth is a small fraction of  the Zeeman shift.   Our data do not follow the constant value theoretically predicted by the dotted line of the hyperfine Paschen-Back description. Instead the ratio increases with  the $\Delta \nu_L,B^H_{\rm Center}$ values.\\
\indent After excluding technical issues, as the shift of the probe position within the magnet at very high fields, we have searched a different explanation. As in Sec. IIA, for the rubidium first resonance line the fine Paschen-Back regime is fully reached for a magnetic field around  500 T. Fig.~\ref{gjratio} explores a range of $B$ values lower than this limit.  Nevertheless we  have calculated the eigenergies of all the fine structure levels, {\it i.e.}, both 5$^2P_{3/2}$ and  5$^2P_{1/2}$ states,  including the hyperfine outdiagonal matrix elements as in ref.~\cite{Arimondo:1977}. The operating magnetic field, "low" for fully reaching the fine Paschen-Back  regime, produces a deviation of the $|5^2P_{3/2}; J_e=3/2, m_{Je}=1/2; I, m_{Ie}\rangle$ energies (those of the $\sigma^+$Li transitions) from the hyperfine Paschen-Back prediction, at the $\approx 10^{-4}$ level. As a consequence also the  $B^H_{\rm Center}/B^L_{\rm Center}$ ratio is modified.\\
\indent As starting point,  the theoretical analysis for the  fine Paschen-Back regime  is based on the orbital Land\'e $g_L$-factor equal 1 and the spin $g_S$-factor equal to $g_{5S}$, leading  within  the Russell-Saunders coupling to the $g_{5P_{3/2}}$ value presented in Section~\ref{dataanalysis}. This analysis, represented in Fig.~\ref{gjratio} by the black dashed line, reproducing the observed behaviour at low magnetic fields,  agrees qualitatively with the measured increase at high magnetic fields.\\
\indent By exploring the role of the $g$-factor values on the high field slope of the $B^H_{\rm Center}/B^L_{\rm Center}$ ratio, we find that the experimental data can be reproduced by modifying $g_L$ and $g_S$.  Several contributions modify those values, as  perturbations by excited core states~\cite{Phillips:1952}, configuration mixing~\cite{ChildsGoodman:1971}, combined action of  exchange core polarization and spin-orbit interaction~\cite{GosselFlambaum:2013}, relativistic and  QED corrections ~\cite{Labzowsky:1999,Goidenko:2002,Indelicato:2007,GosselFlambaum:2013}.  By scanning the $[g_L,g_S]$  plane we reach a good agreement between theory and experiment as shown by the red continuous line in Fig~\ref{gjratio}. The data are fitted by all the values lying on the  line segment bounded by the $[1.0012(2), g_{5S}]$ and $[1.,2.0049(2)]$ points. The first extreme assumes that an atomic perturbation modifies the  $g_L$-factor of the $5P$ state without modification of $g_S$.  The second extreme assumes no perturbation on $g_L$ and a $g_S$ increase larger than the  predicted relativistic and  QED corrections. Because our single result cannot discriminate between all these mechanisms, only an atomic physics  theoretical calculation may determine the precise corrections for the $g$-factors. The $[g_L,g_S]$ combinations lying on the above segment lead to very close values for the $5P_{3/2}$  Land\'e factor globally described by
\begin{equation}
g_{5P_{3/2}}=1.33494(15),
\label{gfactor}
\end{equation}
with included error propagation. This value lies between the one reported in the review~\cite{Arimondo:1977} and that derived in Sec.~\ref{dataanalysis} from our reanalysis of the level crossing of~\cite{BelinSvanberg:1971}. It is more precise than both of them. \\
 \indent Owing to the $5\times10^{-4}$ fractional difference between the above $g_{5P_{3/2}}$ value and the Russel-Saunders starting value, the linear relation between $B_{PU}$ and $B_{H4}$ of Sec.~\ref{sec:setup} is modified by a quantity roughly equal to the the error bar of that relation. \\
\indent

\begin{figure}[t!]
\centering
\includegraphics[width=7 cm]{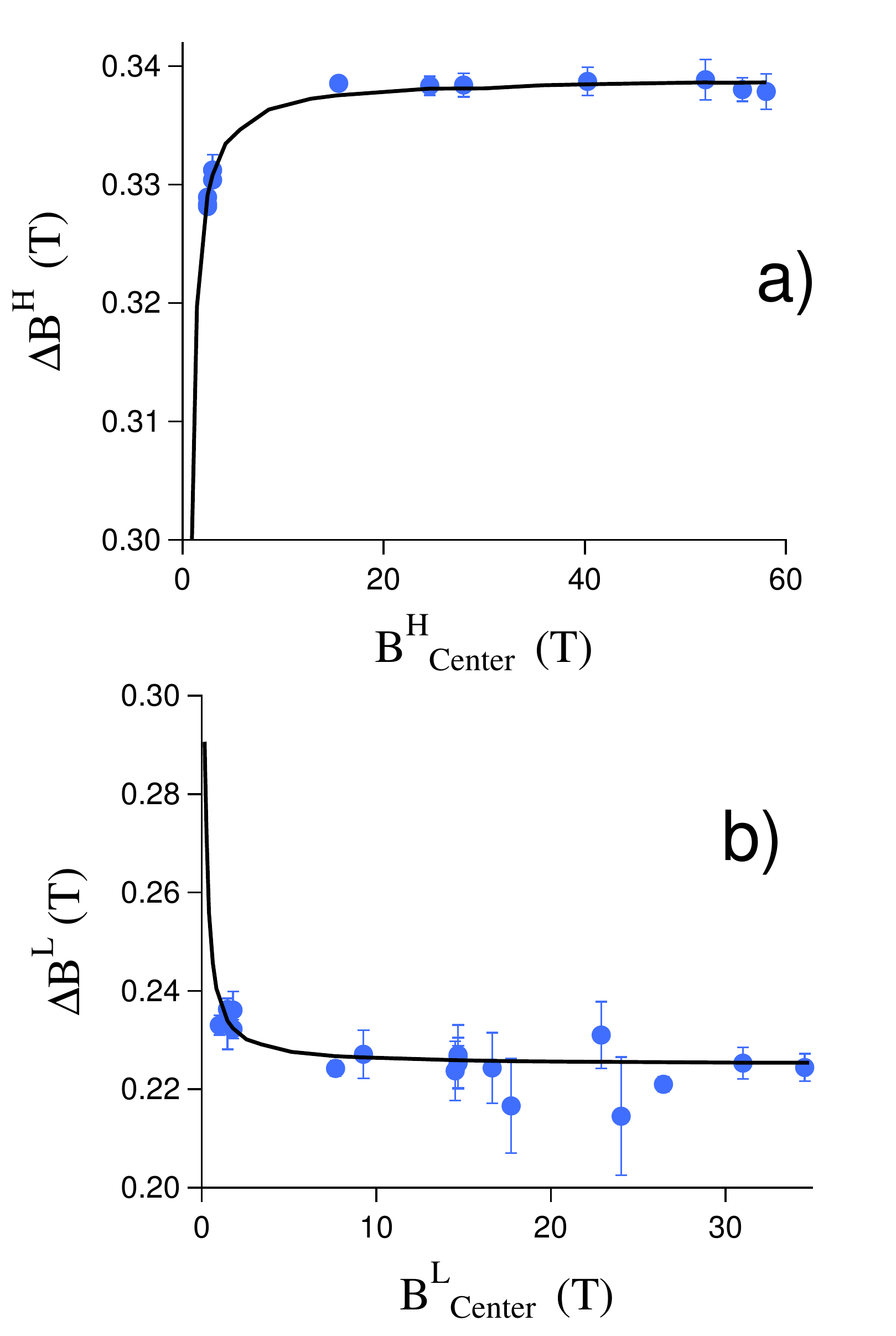}
\caption{Hyperfine interaction splitting  of  the $\sigma^+$Hi  absorption lines, in (a),  and $\sigma^+$Li, in (b), versus the magnetic field center of those lines. Dots for the experimental data with their error bars, and theoretical predictions derived from the Hamiltonan eigenenergies. The constant values at high field correspond to the predictions of the hyperfine/fine  Paschen-Back regime. The $g_{5P_{3/2}}$ value of Eq.~\eqref{gfactor} is inserted into the theoretical analysis. }
\label{NuclearStructure}
\end{figure}
\subsection{Hyperfine structure}
As a test of the comparison between experimental results and theoretical predictions, we have examined the magnetic field separations between the nuclear components of the $\sigma^+$Hi and $\sigma^+$Li lines. We have measured  the $^{87}$Rb quantities $\Delta B^{\mathrm H}= B_{\rm H4} -B_{\mathrm H1}$  and  $\Delta B^{\mathrm L}= B_{\mathrm L1} -B_{\mathrm L4}$ as a function of $BH_{\mathrm Center}$ and $BL_{\mathrm Center}$, respectively. The data and the theoretical predictions are plotted in Fig.~\ref{NuclearStructure}. For the few experimental data having large error bars, the input fiber  Faraday rotation reduces the fluorescence signal intensity and deteriorates the spectra fitting procedure. The nuclear structure is entirely produced by the hyperfine coupling between electronic and nuclear spins, because owing to the $\Delta m_I=0$ selection  the nuclear Zeeman energy does not modify the resonant frequency.  Within the hyperfine Paschen-Back  regime where  $\mathbf{J}$ and $\mathbf{I}$ are fully decoupled,  the $m_I$ levels are nearly equally spaced, producing a constant frequency separation, as shown in  Figs.~\ref{energiesspectra} (b) and (d). That is  not true at the low $B$ fields where $\mathbf{F}$ is the good quantum number.  This difference explains why the theoretical curves rise up or fall down before saturating to a well defined value. The  experimental/theoretical agreement demonstrates the precision of our measurements and the correctness of the atomic eigenenergy derivation. \\
\indent The  hyperfine  Paschen-Back regime is reached for $\mu_B B$ larger than the ground state hyperfine splitting.  For rubidium Sec.~\ref{Paramagnetism} places this transition   around 0.5 T.  Instead the data of Fig.~\ref{NuclearStructure} demonstrate that the transition takes place at different magnetic fields, starting roughly around 0.4 T or 1.5 T depending on the observable, and terminating at higher fields.  The excited state Land\'e $g$-factor determines the magnetic field amplitudes where constant values of $\Delta B^{\mathrm H}$ and $\Delta B^{\mathrm L}$ are reached.  However  the smooth transitions between the different regimes and our error bars do not allow a determination of the $g$-factor. Notice that while the $g$-factor of Eq.~\eqref{gfactor} was used for the theoretical predictions of the figure, the use of the Russell-Saunders produces  theoretical curves modified on their high-field value, but with no visible modification in the transition regions.

\section{Conclusion}
 We have performed high resolution spectroscopy of a  rubidium optical transition at a field slightly larger than 58 T. Our Rb sensor shows performances  more than an order of magnitudes better than standard pick-up coils in terms of uncertainty, compactness and direct access to a micrometer size explored region. \\
 \indent The high precision data collected in a two-week run at the LNCMI facility  allowed an investigation of the  atomic response for high magnetic fields not fully explored previously. We have investigated in detail the transition between different magnetic regimes. We were forced to analyze theoretically our data on the basis of a treatment typically applied only to the fine Paschen-Back regime, even operating at magnetic fields lower than those naively associated to that regime. The focus of our experiment was to test  rubidium atom as a magnetometer, therefore we have not accumulated enough data, as for the Pi lines, and none for the $\sigma^-$ transitions.  A complete investigation will increase the precision of the measured excited state Land\'e $g$-factor.  \\
\indent While our work relies on the Doppler limited absorption spectroscopy, the application of sub-Doppler spectroscopy will lead to an increased resolution by a factor hundred. In our setup the observation of sub-Doppler absorption features relies on technical improvements. Another straightforward way to improve the probe precision is to operate with a cell containing a single rubidium isotope.   \\
\indent Our rubidium magnetometry accuracy is presently limited by the $g_{5P3/2}$ value, whose precision even if  improved by us  cannot yet compete with the  hydrogen nuclear magnetic resonance. When the precision of the $g$-factor  and of  the diamagnetic corrections will be improved,  the Rb magnetometer accuracy could compete with the hydrogen NMR magnetometer.\\
\indent Our results opens the way to dilute matter optical tests in high magnetic fields. Precise measurements of $g$-factors of excited states at a level interesting to verify QED predictions appear feasible. The use  of other atomic transitions or of very narrow optical transitions as in alkaline-earths will expand the atomic physics at high magnetic fields and its applications.

\section*{Acknowledgments}
This research has been partially supported through NEXT (Grant No. ANR-10-LABX-0037) in the framework of the "Programme des Investissements d''Avenir".   EA acknowledges financial support from the Chair d'Excellence Pierre de Fermat of the Conseil Regional Midi-Pyren\'ees. The authors thank S. George, N. Bruyant, J. B\'eard and S. Scotto for technical support, and are grateful to R. Mathevel for very useful suggestions on the  manuscript.

\bibliography{Optical3}

\providecommand{\noopsort}[1]{}\providecommand{\singleletter}[1]{#1}%
\begin{thebibliography}{38}%
\makeatletter
\providecommand \@ifxundefined [1]{%
 \@ifx{#1\undefined}
}%
\providecommand \@ifnum [1]{%
 \ifnum #1\expandafter \@firstoftwo
 \else \expandafter \@secondoftwo
 \fi
}%
\providecommand \@ifx [1]{%
 \ifx #1\expandafter \@firstoftwo
 \else \expandafter \@secondoftwo
 \fi
}%
\providecommand \natexlab [1]{#1}%
\providecommand \enquote  [1]{``#1''}%
\providecommand \bibnamefont  [1]{#1}%
\providecommand \bibfnamefont [1]{#1}%
\providecommand \citenamefont [1]{#1}%
\providecommand \href@noop [0]{\@secondoftwo}%
\providecommand \href [0]{\begingroup \@sanitize@url \@href}%
\providecommand \@href[1]{\@@startlink{#1}\@@href}%
\providecommand \@@href[1]{\endgroup#1\@@endlink}%
\providecommand \@sanitize@url [0]{\catcode `\\12\catcode `\$12\catcode
  `\&12\catcode `\#12\catcode `\^12\catcode `\_12\catcode `\%12\relax}%
\providecommand \@@startlink[1]{}%
\providecommand \@@endlink[0]{}%
\providecommand \url  [0]{\begingroup\@sanitize@url \@url }%
\providecommand \@url [1]{\endgroup\@href {#1}{\urlprefix }}%
\providecommand \urlprefix  [0]{URL }%
\providecommand \Eprint [0]{\href }%
\providecommand \doibase [0]{http://dx.doi.org/}%
\providecommand \selectlanguage [0]{\@gobble}%
\providecommand \bibinfo  [0]{\@secondoftwo}%
\providecommand \bibfield  [0]{\@secondoftwo}%
\providecommand \translation [1]{[#1]}%
\providecommand \BibitemOpen [0]{}%
\providecommand \bibitemStop [0]{}%
\providecommand \bibitemNoStop [0]{.\EOS\space}%
\providecommand \EOS [0]{\spacefactor3000\relax}%
\providecommand \BibitemShut  [1]{\csname bibitem#1\endcsname}%
\let\auto@bib@innerbib\@empty
\bibitem [{\citenamefont {Kominis}\ \emph {et~al.}(2003)\citenamefont
  {Kominis}, \citenamefont {Kornack}, \citenamefont {Allred},\ and\
  \citenamefont {Romalis}}]{KominisRomalis:2003}%
  \BibitemOpen
  \bibfield  {author} {\bibinfo {author} {\bibfnamefont {I.~K.}\ \bibnamefont
  {Kominis}}, \bibinfo {author} {\bibfnamefont {T.~W.}\ \bibnamefont
  {Kornack}}, \bibinfo {author} {\bibfnamefont {J.~C.}\ \bibnamefont {Allred}},
  \ and\ \bibinfo {author} {\bibfnamefont {M.~V.}\ \bibnamefont {Romalis}},\
  }\href@noop {} {\bibfield  {journal} {\bibinfo  {journal} {Nature}\ }\textbf
  {\bibinfo {volume} {422}},\ \bibinfo {pages} {596} (\bibinfo {year}
  {2003})}\BibitemShut {NoStop}%
\bibitem [{\citenamefont {Vengalattore}\ \emph {et~al.}(2007)\citenamefont
  {Vengalattore}, \citenamefont {Higbie}, \citenamefont {Leslie}, \citenamefont
  {Guzman}, \citenamefont {Sadler},\ and\ \citenamefont
  {Stamper-Kurn}}]{VengalattoreStamperKurn:2007}%
  \BibitemOpen
  \bibfield  {author} {\bibinfo {author} {\bibfnamefont {M.}~\bibnamefont
  {Vengalattore}}, \bibinfo {author} {\bibfnamefont {J.~M.}\ \bibnamefont
  {Higbie}}, \bibinfo {author} {\bibfnamefont {S.~R.}\ \bibnamefont {Leslie}},
  \bibinfo {author} {\bibfnamefont {J.}~\bibnamefont {Guzman}}, \bibinfo
  {author} {\bibfnamefont {L.~E.}\ \bibnamefont {Sadler}}, \ and\ \bibinfo
  {author} {\bibfnamefont {D.~M.}\ \bibnamefont {Stamper-Kurn}},\ }\href
  {\doibase 10.1103/PhysRevLett.98.200801} {\bibfield  {journal} {\bibinfo
  {journal} {Phys. Rev. Lett.}\ }\textbf {\bibinfo {volume} {98}},\ \bibinfo
  {pages} {200801} (\bibinfo {year} {2007})}\BibitemShut {NoStop}%
\bibitem [{\citenamefont {Mamin}\ \emph {et~al.}(2007)\citenamefont {Mamin},
  \citenamefont {Poggio}, \citenamefont {Degen},\ and\ \citenamefont
  {Rugar}}]{MaminRugar:2007}%
  \BibitemOpen
  \bibfield  {author} {\bibinfo {author} {\bibfnamefont {H.~J.}\ \bibnamefont
  {Mamin}}, \bibinfo {author} {\bibfnamefont {M.}~\bibnamefont {Poggio}},
  \bibinfo {author} {\bibfnamefont {C.~L.}\ \bibnamefont {Degen}}, \ and\
  \bibinfo {author} {\bibfnamefont {D.}~\bibnamefont {Rugar}},\ }\href@noop {}
  {\bibfield  {journal} {\bibinfo  {journal} {Nat Nano}\ }\textbf {\bibinfo
  {volume} {2}},\ \bibinfo {pages} {301} (\bibinfo {year} {2007})}\BibitemShut
  {NoStop}%
\bibitem [{\citenamefont {Taylor}\ \emph {et~al.}(2008)\citenamefont {Taylor},
  \citenamefont {Cappellaro}, \citenamefont {Childress}, \citenamefont {Jiang},
  \citenamefont {Budker}, \citenamefont {Hemmer}, \citenamefont {Yacoby},
  \citenamefont {Walsworth},\ and\ \citenamefont {Lukin}}]{TaylorLukin:2008}%
  \BibitemOpen
  \bibfield  {author} {\bibinfo {author} {\bibfnamefont {J.~M.}\ \bibnamefont
  {Taylor}}, \bibinfo {author} {\bibfnamefont {P.}~\bibnamefont {Cappellaro}},
  \bibinfo {author} {\bibfnamefont {L.}~\bibnamefont {Childress}}, \bibinfo
  {author} {\bibfnamefont {L.}~\bibnamefont {Jiang}}, \bibinfo {author}
  {\bibfnamefont {D.}~\bibnamefont {Budker}}, \bibinfo {author} {\bibfnamefont
  {P.~R.}\ \bibnamefont {Hemmer}}, \bibinfo {author} {\bibfnamefont
  {A.}~\bibnamefont {Yacoby}}, \bibinfo {author} {\bibfnamefont
  {R.}~\bibnamefont {Walsworth}}, \ and\ \bibinfo {author} {\bibfnamefont
  {M.~D.}\ \bibnamefont {Lukin}},\ }\href@noop {} {\bibfield  {journal}
  {\bibinfo  {journal} {Nat. Phys.}\ }\textbf {\bibinfo {volume} {4}},\
  \bibinfo {pages} {810} (\bibinfo {year} {2008})}\BibitemShut {NoStop}%
\bibitem [{\citenamefont {Pham}\ \emph {et~al.}(2011)\citenamefont {Pham},
  \citenamefont {Sage}, \citenamefont {Stanwix}, \citenamefont {Yeung},
  \citenamefont {Glenn}, \citenamefont {Trifonov}, \citenamefont {Cappellaro},
  \citenamefont {Hemmer}, \citenamefont {Lukin}, \citenamefont {Park},
  \citenamefont {Yacoby},\ and\ \citenamefont {Walsworth}}]{PhamWalsorth:2011}%
  \BibitemOpen
  \bibfield  {author} {\bibinfo {author} {\bibfnamefont {L.~M.}\ \bibnamefont
  {Pham}}, \bibinfo {author} {\bibfnamefont {D.~L.}\ \bibnamefont {Sage}},
  \bibinfo {author} {\bibfnamefont {P.~L.}\ \bibnamefont {Stanwix}}, \bibinfo
  {author} {\bibfnamefont {T.~K.}\ \bibnamefont {Yeung}}, \bibinfo {author}
  {\bibfnamefont {D.}~\bibnamefont {Glenn}}, \bibinfo {author} {\bibfnamefont
  {A.}~\bibnamefont {Trifonov}}, \bibinfo {author} {\bibfnamefont
  {P.}~\bibnamefont {Cappellaro}}, \bibinfo {author} {\bibfnamefont {P.~R.}\
  \bibnamefont {Hemmer}}, \bibinfo {author} {\bibfnamefont {M.~D.}\
  \bibnamefont {Lukin}}, \bibinfo {author} {\bibfnamefont {H.}~\bibnamefont
  {Park}}, \bibinfo {author} {\bibfnamefont {A.}~\bibnamefont {Yacoby}}, \ and\
  \bibinfo {author} {\bibfnamefont {R.~L.}\ \bibnamefont {Walsworth}},\ }\href
  {http://stacks.iop.org/1367-2630/13/i=4/a=045021} {\bibfield  {journal}
  {\bibinfo  {journal} {NJP}\ }\textbf {\bibinfo {volume} {13}},\ \bibinfo
  {pages} {045021} (\bibinfo {year} {2011})}\BibitemShut {NoStop}%
\bibitem [{\citenamefont {Rondin}\ \emph {et~al.}(2014)\citenamefont {Rondin},
  \citenamefont {Tetienne}, \citenamefont {Hingant}, \citenamefont {Roch},
  \citenamefont {Maletinsky},\ and\ \citenamefont
  {Jacques}}]{RondinJacques:2014}%
  \BibitemOpen
  \bibfield  {author} {\bibinfo {author} {\bibfnamefont {L.}~\bibnamefont
  {Rondin}}, \bibinfo {author} {\bibfnamefont {J.-P.}\ \bibnamefont
  {Tetienne}}, \bibinfo {author} {\bibfnamefont {T.}~\bibnamefont {Hingant}},
  \bibinfo {author} {\bibfnamefont {J.-F.}\ \bibnamefont {Roch}}, \bibinfo
  {author} {\bibfnamefont {P.}~\bibnamefont {Maletinsky}}, \ and\ \bibinfo
  {author} {\bibfnamefont {V.}~\bibnamefont {Jacques}},\ }\href@noop {}
  {\bibfield  {journal} {\bibinfo  {journal} {Rep. Progr. Phys.}\ }\textbf
  {\bibinfo {volume} {77}},\ \bibinfo {pages} {056503} (\bibinfo {year}
  {2014})}\BibitemShut {NoStop}%
\bibitem [{\citenamefont {Budker}\ and\ \citenamefont
  {Kimball}(2013)}]{Budker:2013}%
  \BibitemOpen
  \bibfield  {author} {\bibinfo {author} {\bibfnamefont {F.}~\bibnamefont
  {Budker}, \bibfnamefont {D.~Derek}}\ and\ \bibinfo {author} {\bibfnamefont
  {J.}~\bibnamefont {Kimball}},\ }\href@noop {} {\emph {\bibinfo {title}
  {Optical Magnetometry}}}\ (\bibinfo  {publisher} {Cambridge University
  Press},\ \bibinfo {year} {2013})\BibitemShut {NoStop}%
\bibitem [{\citenamefont {Garn}\ \emph {et~al.}(1966)\citenamefont {Garn},
  \citenamefont {Caird}, \citenamefont {Thomson},\ and\ \citenamefont
  {Fowler}}]{Garn:1966}%
  \BibitemOpen
  \bibfield  {author} {\bibinfo {author} {\bibfnamefont {W.~B.}\ \bibnamefont
  {Garn}}, \bibinfo {author} {\bibfnamefont {R.~S.}\ \bibnamefont {Caird}},
  \bibinfo {author} {\bibfnamefont {D.~B.}\ \bibnamefont {Thomson}}, \ and\
  \bibinfo {author} {\bibfnamefont {C.~M.}\ \bibnamefont {Fowler}},\
  }\href@noop {} {\bibfield  {journal} {\bibinfo  {journal} {Rev. Sci.
  Instrum.}\ }\textbf {\bibinfo {volume} {37}},\ \bibinfo {pages} {762}
  (\bibinfo {year} {1966})}\BibitemShut {NoStop}%
\bibitem [{\citenamefont {Hori}\ \emph {et~al.}(1982)\citenamefont {Hori},
  \citenamefont {Miki},\ and\ \citenamefont {Date}}]{Hori:1982}%
  \BibitemOpen
  \bibfield  {author} {\bibinfo {author} {\bibfnamefont {H.}~\bibnamefont
  {Hori}}, \bibinfo {author} {\bibfnamefont {M.}~\bibnamefont {Miki}}, \ and\
  \bibinfo {author} {\bibfnamefont {M.}~\bibnamefont {Date}},\ }\href@noop {}
  {\bibfield  {journal} {\bibinfo  {journal} {J. Phys. Soc. Japan}\ }\textbf
  {\bibinfo {volume} {51}},\ \bibinfo {pages} {1566} (\bibinfo {year}
  {1982})}\BibitemShut {NoStop}%
\bibitem [{\citenamefont {Gomez}\ \emph {et~al.}(2014)\citenamefont {Gomez},
  \citenamefont {Hansen}, \citenamefont {Peterson}, \citenamefont {Bliss},
  \citenamefont {Carlson}, \citenamefont {Lamppa}, \citenamefont {Schroen},\
  and\ \citenamefont {Rochau}}]{Gomez:2014}%
  \BibitemOpen
  \bibfield  {author} {\bibinfo {author} {\bibfnamefont {M.~R.}\ \bibnamefont
  {Gomez}}, \bibinfo {author} {\bibfnamefont {S.~B.}\ \bibnamefont {Hansen}},
  \bibinfo {author} {\bibfnamefont {K.~J.}\ \bibnamefont {Peterson}}, \bibinfo
  {author} {\bibfnamefont {D.~E.}\ \bibnamefont {Bliss}}, \bibinfo {author}
  {\bibfnamefont {A.~L.}\ \bibnamefont {Carlson}}, \bibinfo {author}
  {\bibfnamefont {D.~C.}\ \bibnamefont {Lamppa}}, \bibinfo {author}
  {\bibfnamefont {D.~G.}\ \bibnamefont {Schroen}}, \ and\ \bibinfo {author}
  {\bibfnamefont {G.~A.}\ \bibnamefont {Rochau}},\ }\href@noop {} {\bibfield
  {journal} {\bibinfo  {journal} {Rev. Sci. Instrum.}\ }\textbf {\bibinfo
  {volume} {85}},\ \bibinfo {pages} {11E609} (\bibinfo {year}
  {2014})}\BibitemShut {NoStop}%
\bibitem [{\citenamefont {Banasek}\ \emph {et~al.}(2016)\citenamefont
  {Banasek}, \citenamefont {Engelbrecht}, \citenamefont {Pikuz}, \citenamefont
  {Shelkovenko},\ and\ \citenamefont {Hammer}}]{Banasek:2016}%
  \BibitemOpen
  \bibfield  {author} {\bibinfo {author} {\bibfnamefont {J.~T.}\ \bibnamefont
  {Banasek}}, \bibinfo {author} {\bibfnamefont {J.~T.}\ \bibnamefont
  {Engelbrecht}}, \bibinfo {author} {\bibfnamefont {S.~A.}\ \bibnamefont
  {Pikuz}}, \bibinfo {author} {\bibfnamefont {T.~A.}\ \bibnamefont
  {Shelkovenko}}, \ and\ \bibinfo {author} {\bibfnamefont {D.~A.}\ \bibnamefont
  {Hammer}},\ }\href@noop {} {\bibfield  {journal} {\bibinfo  {journal} {Rev.
  Sci. Instrum.}\ }\textbf {\bibinfo {volume} {87}},\ \bibinfo {pages} {103506}
  (\bibinfo {year} {2016})}\BibitemShut {NoStop}%
\bibitem [{\citenamefont {George}\ \emph {et~al.}(2017)\citenamefont {George},
  \citenamefont {Bruyant}, \citenamefont {B\'eard}, \citenamefont {Scotto},
  \citenamefont {Arimondo}, \citenamefont {Battesti}, \citenamefont
  {Ciampini},\ and\ \citenamefont {Rizzo}}]{George:2017}%
  \BibitemOpen
  \bibfield  {author} {\bibinfo {author} {\bibfnamefont {S.}~\bibnamefont
  {George}}, \bibinfo {author} {\bibfnamefont {N.}~\bibnamefont {Bruyant}},
  \bibinfo {author} {\bibfnamefont {J.~B.}\ \bibnamefont {B\'eard}}, \bibinfo
  {author} {\bibfnamefont {S.}~\bibnamefont {Scotto}}, \bibinfo {author}
  {\bibfnamefont {E.}~\bibnamefont {Arimondo}}, \bibinfo {author}
  {\bibfnamefont {R.}~\bibnamefont {Battesti}}, \bibinfo {author}
  {\bibfnamefont {D.}~\bibnamefont {Ciampini}}, \ and\ \bibinfo {author}
  {\bibfnamefont {C.}~\bibnamefont {Rizzo}},\ }\href@noop {} {\bibfield
  {journal} {\bibinfo  {journal} {Rev. Sci. Instrum.}\ }\textbf {\bibinfo
  {volume} {88}},\ \bibinfo {pages} {073102} (\bibinfo {year}
  {2017})}\BibitemShut {NoStop}%
\bibitem [{\citenamefont {Kopfermann}(1958)}]{Kopfermann:1958}%
  \BibitemOpen
  \bibfield  {author} {\bibinfo {author} {\bibfnamefont {H.}~\bibnamefont
  {Kopfermann}},\ }\href@noop {} {\emph {\bibinfo {title} {Nuclear Moments}}}\
  (\bibinfo  {publisher} {Academic Press, New York},\ \bibinfo {year}
  {1958})\BibitemShut {NoStop}%
\bibitem [{\citenamefont {Lipson}\ \emph {et~al.}()\citenamefont {Lipson},
  \citenamefont {Fletcher},\ and\ \citenamefont {Larson}}]{LipsonLarson:1986}%
  \BibitemOpen
  \bibfield  {author} {\bibinfo {author} {\bibfnamefont {S.~J.}\ \bibnamefont
  {Lipson}}, \bibinfo {author} {\bibfnamefont {G.~D.}\ \bibnamefont
  {Fletcher}}, \ and\ \bibinfo {author} {\bibfnamefont {D.~J.}\ \bibnamefont
  {Larson}},\ }\href@noop {} {\bibfield  {journal} {\bibinfo  {journal} {Phys.
  Rev. Lett.}\ }\textbf {\bibinfo {volume} {57}},\ \bibinfo {pages}
  {567}}\BibitemShut {NoStop}%
\bibitem [{\citenamefont {Fortson}()}]{Fortson:1987}%
  \BibitemOpen
  \bibfield  {author} {\bibinfo {author} {\bibfnamefont {N.}~\bibnamefont
  {Fortson}},\ }\href@noop {} {\bibfield  {journal} {\bibinfo  {journal} {Phys.
  Rev. Lett.}\ }\textbf {\bibinfo {volume} {59}},\ \bibinfo {pages}
  {988}}\BibitemShut {NoStop}%
\bibitem [{\citenamefont {Arimondo}\ \emph {et~al.}(1977)\citenamefont
  {Arimondo}, \citenamefont {Inguscio},\ and\ \citenamefont
  {Violino}}]{Arimondo:1977}%
  \BibitemOpen
  \bibfield  {author} {\bibinfo {author} {\bibfnamefont {E.}~\bibnamefont
  {Arimondo}}, \bibinfo {author} {\bibfnamefont {M.}~\bibnamefont {Inguscio}},
  \ and\ \bibinfo {author} {\bibfnamefont {P.}~\bibnamefont {Violino}},\
  }\href@noop {} {\bibfield  {journal} {\bibinfo  {journal} {Rev. Mod. Phys.}\
  }\textbf {\bibinfo {volume} {49}},\ \bibinfo {pages} {31} (\bibinfo {year}
  {1977})}\BibitemShut {NoStop}%
\bibitem [{\citenamefont {Foot}(2012)}]{Foot:2012}%
  \BibitemOpen
  \bibfield  {author} {\bibinfo {author} {\bibfnamefont {C.}~\bibnamefont
  {Foot}},\ }\href@noop {} {\emph {\bibinfo {title} {Atomic Physics}}}\
  (\bibinfo  {publisher} {Oxford University Press},\ \bibinfo {year}
  {2012})\BibitemShut {NoStop}%
\bibitem [{\citenamefont {Schiff}\ and\ \citenamefont
  {Snyder}(1939)}]{SchiffSnyder:1939}%
  \BibitemOpen
  \bibfield  {author} {\bibinfo {author} {\bibfnamefont {L.}~\bibnamefont
  {Schiff}}\ and\ \bibinfo {author} {\bibfnamefont {H.}~\bibnamefont
  {Snyder}},\ }\href@noop {} {\bibfield  {journal} {\bibinfo  {journal} {Phys.
  Rev.}\ }\textbf {\bibinfo {volume} {55}},\ \bibinfo {pages} {59} (\bibinfo
  {year} {1939})}\BibitemShut {NoStop}%
\bibitem [{\citenamefont {Garstang}(1977)}]{Garstang:1977}%
  \BibitemOpen
  \bibfield  {author} {\bibinfo {author} {\bibfnamefont {R.}~\bibnamefont
  {Garstang}},\ }\href@noop {} {\bibfield  {journal} {\bibinfo  {journal} {Rep.
  Progr. Phys.}\ }\textbf {\bibinfo {volume} {40}},\ \bibinfo {pages} {105}
  (\bibinfo {year} {1977})}\BibitemShut {NoStop}%
\bibitem [{\citenamefont {Otto}\ \emph {et~al.}(2002)\citenamefont {Otto},
  \citenamefont {Gamperling}, \citenamefont {Hofacker}, \citenamefont {Meyer},
  \citenamefont {Pagliari}, \citenamefont {Stifter}, \citenamefont {Krauss},\
  and\ \citenamefont {H\"uttner}}]{Otto:2002}%
  \BibitemOpen
  \bibfield  {author} {\bibinfo {author} {\bibfnamefont {P.}~\bibnamefont
  {Otto}}, \bibinfo {author} {\bibfnamefont {M.}~\bibnamefont {Gamperling}},
  \bibinfo {author} {\bibfnamefont {M.}~\bibnamefont {Hofacker}}, \bibinfo
  {author} {\bibfnamefont {T.}~\bibnamefont {Meyer}}, \bibinfo {author}
  {\bibfnamefont {V.}~\bibnamefont {Pagliari}}, \bibinfo {author}
  {\bibfnamefont {A.}~\bibnamefont {Stifter}}, \bibinfo {author} {\bibfnamefont
  {M.}~\bibnamefont {Krauss}}, \ and\ \bibinfo {author} {\bibfnamefont
  {W.}~\bibnamefont {H\"uttner}},\ }\href@noop {} {\bibfield  {journal}
  {\bibinfo  {journal} {Chem. Phys.}\ }\textbf {\bibinfo {volume} {282}},\
  \bibinfo {pages} {289} (\bibinfo {year} {2002})}\BibitemShut {NoStop}%
\bibitem [{\citenamefont {Jenkins}\ and\ \citenamefont
  {Segr\`e}(1939)}]{JenkinsSegre:1939}%
  \BibitemOpen
  \bibfield  {author} {\bibinfo {author} {\bibfnamefont {F.}~\bibnamefont
  {Jenkins}}\ and\ \bibinfo {author} {\bibfnamefont {E.}~\bibnamefont
  {Segr\`e}},\ }\href@noop {} {\bibfield  {journal} {\bibinfo  {journal} {Phys.
  Rev.}\ }\textbf {\bibinfo {volume} {55}},\ \bibinfo {pages} {52} (\bibinfo
  {year} {1939})}\BibitemShut {NoStop}%
\bibitem [{\citenamefont {Harting}\ and\ \citenamefont
  {Klinkenberg}(1949)}]{HartingKlinkenberg:1949}%
  \BibitemOpen
  \bibfield  {author} {\bibinfo {author} {\bibfnamefont {D.}~\bibnamefont
  {Harting}}\ and\ \bibinfo {author} {\bibfnamefont {P.}~\bibnamefont
  {Klinkenberg}},\ }\href@noop {} {\bibfield  {journal} {\bibinfo  {journal}
  {Physica}\ }\textbf {\bibinfo {volume} {14}},\ \bibinfo {pages} {669}
  (\bibinfo {year} {1949})}\BibitemShut {NoStop}%
\bibitem [{\citenamefont {Harper}\ and\ \citenamefont
  {Levenson}(1977)}]{HarperLevenson:1977}%
  \BibitemOpen
  \bibfield  {author} {\bibinfo {author} {\bibfnamefont {C.}~\bibnamefont
  {Harper}}\ and\ \bibinfo {author} {\bibfnamefont {M.}~\bibnamefont
  {Levenson}},\ }\href@noop {} {\bibfield  {journal} {\bibinfo  {journal} {Opt.
  Commun.}\ }\textbf {\bibinfo {volume} {20}},\ \bibinfo {pages} {107}
  (\bibinfo {year} {1977})}\BibitemShut {NoStop}%
\bibitem [{\citenamefont {H\"uttner}\ \emph {et~al.}(1996)\citenamefont
  {H\"uttner}, \citenamefont {Otto},\ and\ \citenamefont
  {Gamperling}}]{Huettner:1996}%
  \BibitemOpen
  \bibfield  {author} {\bibinfo {author} {\bibfnamefont {W.}~\bibnamefont
  {H\"uttner}}, \bibinfo {author} {\bibfnamefont {P.}~\bibnamefont {Otto}}, \
  and\ \bibinfo {author} {\bibfnamefont {M.}~\bibnamefont {Gamperling}},\
  }\href@noop {} {\bibfield  {journal} {\bibinfo  {journal} {Phys. Rev. A}\
  }\textbf {\bibinfo {volume} {54}},\ \bibinfo {pages} {1318} (\bibinfo {year}
  {1996})}\BibitemShut {NoStop}%
\bibitem [{\citenamefont {Gallagher}(1994)}]{Gallagher:1994}%
  \BibitemOpen
  \bibfield  {author} {\bibinfo {author} {\bibfnamefont {T.}~\bibnamefont
  {Gallagher}},\ }\href@noop {} {\emph {\bibinfo {title} {Rydberg Atoms}}}\
  (\bibinfo  {publisher} {Cambridge University Press},\ \bibinfo {year}
  {1994})\BibitemShut {NoStop}%
\bibitem [{\citenamefont {Li}\ \emph {et~al.}(2003)\citenamefont {Li},
  \citenamefont {Mourachko}, \citenamefont {Noel},\ and\ \citenamefont
  {Gallagher}}]{LiGallagher:2003}%
  \BibitemOpen
  \bibfield  {author} {\bibinfo {author} {\bibfnamefont {W.}~\bibnamefont
  {Li}}, \bibinfo {author} {\bibfnamefont {I.}~\bibnamefont {Mourachko}},
  \bibinfo {author} {\bibfnamefont {M.}~\bibnamefont {Noel}}, \ and\ \bibinfo
  {author} {\bibfnamefont {T.}~\bibnamefont {Gallagher}},\ }\href@noop {}
  {\bibfield  {journal} {\bibinfo  {journal} {Phys. Rev. A}\ }\textbf {\bibinfo
  {volume} {67}},\ \bibinfo {pages} {052502} (\bibinfo {year}
  {2003})}\BibitemShut {NoStop}%
\bibitem [{\citenamefont {Steck}(2001)}]{Steck87:2001}%
  \BibitemOpen
  \bibfield  {author} {\bibinfo {author} {\bibfnamefont {D.}~\bibnamefont
  {Steck}},\ }\href@noop {} {\enquote {\bibinfo {title} {Rubidium 87 {D} line
  data, 2001},}\ }\bibinfo {howpublished}
  {$george.ph.utexas.edu/~dsteck/alkalidata$ $rubidium87numbers.pdf$} (\bibinfo
  {year} {2001})\BibitemShut {NoStop}%
\bibitem [{\citenamefont {Tiedeman}\ and\ \citenamefont
  {Robinson}(1977)}]{Tiedeman:1977}%
  \BibitemOpen
  \bibfield  {author} {\bibinfo {author} {\bibfnamefont {J.~S.}\ \bibnamefont
  {Tiedeman}}\ and\ \bibinfo {author} {\bibfnamefont {H.~G.}\ \bibnamefont
  {Robinson}},\ }\href@noop {} {\bibfield  {journal} {\bibinfo  {journal}
  {Phys. Rev. Lett.}\ }\textbf {\bibinfo {volume} {39}},\ \bibinfo {pages}
  {602} (\bibinfo {year} {1977})}\BibitemShut {NoStop}%
\bibitem [{\citenamefont {Mohr}\ \emph {et~al.}(2016)\citenamefont {Mohr},
  \citenamefont {Newell},\ and\ \citenamefont {Taylor}}]{CODATA:2016}%
  \BibitemOpen
  \bibfield  {author} {\bibinfo {author} {\bibfnamefont {P.~J.}\ \bibnamefont
  {Mohr}}, \bibinfo {author} {\bibfnamefont {D.~B.}\ \bibnamefont {Newell}}, \
  and\ \bibinfo {author} {\bibfnamefont {B.~N.}\ \bibnamefont {Taylor}},\
  }\href@noop {} {\bibfield  {journal} {\bibinfo  {journal} {Rev. Mod. Phys.}\
  }\textbf {\bibinfo {volume} {88}},\ \bibinfo {pages} {035009} (\bibinfo
  {year} {2016})}\BibitemShut {NoStop}%
\bibitem [{\citenamefont {Belin}\ and\ \citenamefont
  {Svanberg}(1971)}]{BelinSvanberg:1971}%
  \BibitemOpen
  \bibfield  {author} {\bibinfo {author} {\bibfnamefont {G.}~\bibnamefont
  {Belin}}\ and\ \bibinfo {author} {\bibfnamefont {S.}~\bibnamefont
  {Svanberg}},\ }\href@noop {} {\bibfield  {journal} {\bibinfo  {journal}
  {Physica Scripta}\ }\textbf {\bibinfo {volume} {4}},\ \bibinfo {pages} {269}
  (\bibinfo {year} {1971})}\BibitemShut {NoStop}%
\bibitem [{\citenamefont {Ye}\ \emph {et~al.}(1996)\citenamefont {Ye},
  \citenamefont {Swartz}, \citenamefont {Jungner},\ and\ \citenamefont
  {Hall}}]{YeHall:1996}%
  \BibitemOpen
  \bibfield  {author} {\bibinfo {author} {\bibfnamefont {J.}~\bibnamefont
  {Ye}}, \bibinfo {author} {\bibfnamefont {S.}~\bibnamefont {Swartz}}, \bibinfo
  {author} {\bibfnamefont {P.}~\bibnamefont {Jungner}}, \ and\ \bibinfo
  {author} {\bibfnamefont {J.~L.}\ \bibnamefont {Hall}},\ }\href@noop {}
  {\bibfield  {journal} {\bibinfo  {journal} {Opt. Lett.}\ }\textbf {\bibinfo
  {volume} {21}},\ \bibinfo {pages} {1280} (\bibinfo {year}
  {1996})}\BibitemShut {NoStop}%
\bibitem [{\citenamefont {Labzowsky}\ \emph {et~al.}(1999)\citenamefont
  {Labzowsky}, \citenamefont {Goidenko},\ and\ \citenamefont
  {Pyykk{\"o}}}]{Labzowsky:1999}%
  \BibitemOpen
  \bibfield  {author} {\bibinfo {author} {\bibfnamefont {L.}~\bibnamefont
  {Labzowsky}}, \bibinfo {author} {\bibfnamefont {I.}~\bibnamefont {Goidenko}},
  \ and\ \bibinfo {author} {\bibfnamefont {P.}~\bibnamefont {Pyykk{\"o}}},\
  }\href@noop {} {\bibfield  {journal} {\bibinfo  {journal} {Phys. Lett. A}\
  }\textbf {\bibinfo {volume} {258}},\ \bibinfo {pages} {31 } (\bibinfo {year}
  {1999})}\BibitemShut {NoStop}%
\bibitem [{\citenamefont {Goidenko}\ \emph {et~al.}(2002)\citenamefont
  {Goidenko}, \citenamefont {Labzowsky}, \citenamefont {Plunien},\ and\
  \citenamefont {Soff}}]{Goidenko:2002}%
  \BibitemOpen
  \bibfield  {author} {\bibinfo {author} {\bibfnamefont {I.}~\bibnamefont
  {Goidenko}}, \bibinfo {author} {\bibfnamefont {L.}~\bibnamefont {Labzowsky}},
  \bibinfo {author} {\bibfnamefont {G.}~\bibnamefont {Plunien}}, \ and\
  \bibinfo {author} {\bibfnamefont {G.}~\bibnamefont {Soff}},\ }\href@noop {}
  {\bibfield  {journal} {\bibinfo  {journal} {Phys. Rev. A}\ }\textbf {\bibinfo
  {volume} {66}},\ \bibinfo {pages} {032115} (\bibinfo {year}
  {2002})}\BibitemShut {NoStop}%
\bibitem [{\citenamefont {Indelicato}\ \emph {et~al.}(2007)\citenamefont
  {Indelicato}, \citenamefont {Santos}, \citenamefont {Boucard},\ and\
  \citenamefont {Desclaux}}]{Indelicato:2007}%
  \BibitemOpen
  \bibfield  {author} {\bibinfo {author} {\bibfnamefont {P.}~\bibnamefont
  {Indelicato}}, \bibinfo {author} {\bibfnamefont {J.~P.}\ \bibnamefont
  {Santos}}, \bibinfo {author} {\bibfnamefont {S.}~\bibnamefont {Boucard}}, \
  and\ \bibinfo {author} {\bibfnamefont {J.-P.}\ \bibnamefont {Desclaux}},\
  }\href@noop {} {\bibfield  {journal} {\bibinfo  {journal} {Eur. Phys. J. D}\
  }\textbf {\bibinfo {volume} {45}},\ \bibinfo {pages} {155} (\bibinfo {year}
  {2007})}\BibitemShut {NoStop}%
\bibitem [{\citenamefont {Gossel}\ \emph {et~al.}(2013)\citenamefont {Gossel},
  \citenamefont {Dzuba},\ and\ \citenamefont {Flambaum}}]{GosselFlambaum:2013}%
  \BibitemOpen
  \bibfield  {author} {\bibinfo {author} {\bibfnamefont {G.~H.}\ \bibnamefont
  {Gossel}}, \bibinfo {author} {\bibfnamefont {V.~A.}\ \bibnamefont {Dzuba}}, \
  and\ \bibinfo {author} {\bibfnamefont {V.~V.}\ \bibnamefont {Flambaum}},\
  }\href@noop {} {\bibfield  {journal} {\bibinfo  {journal} {Phys. Rev. A}\
  }\textbf {\bibinfo {volume} {88}},\ \bibinfo {pages} {034501} (\bibinfo
  {year} {2013})}\BibitemShut {NoStop}%
\bibitem [{\citenamefont {Debray}\ and\ \citenamefont
  {Frings}(2013)}]{Debray:2013}%
  \BibitemOpen
  \bibfield  {author} {\bibinfo {author} {\bibfnamefont {F.}~\bibnamefont
  {Debray}}\ and\ \bibinfo {author} {\bibfnamefont {P.}~\bibnamefont
  {Frings}},\ }\href@noop {} {\bibfield  {journal} {\bibinfo  {journal} {C. R.
  Phys.}\ }\textbf {\bibinfo {volume} {14}},\ \bibinfo {pages} {14} (\bibinfo
  {year} {2013})}\BibitemShut {NoStop}%
\bibitem [{\citenamefont {Phillips}(1952)}]{Phillips:1952}%
  \BibitemOpen
  \bibfield  {author} {\bibinfo {author} {\bibfnamefont {M.}~\bibnamefont
  {Phillips}},\ }\href@noop {} {\bibfield  {journal} {\bibinfo  {journal}
  {Phys. Rev.}\ }\textbf {\bibinfo {volume} {88}},\ \bibinfo {pages} {202}
  (\bibinfo {year} {1952})}\BibitemShut {NoStop}%
\bibitem [{\citenamefont {Childs}\ and\ \citenamefont
  {Goodman}(1971)}]{ChildsGoodman:1971}%
  \BibitemOpen
  \bibfield  {author} {\bibinfo {author} {\bibfnamefont {W.~J.}\ \bibnamefont
  {Childs}}\ and\ \bibinfo {author} {\bibfnamefont {L.~S.}\ \bibnamefont
  {Goodman}},\ }\href@noop {} {\bibfield  {journal} {\bibinfo  {journal} {Phys.
  Rev. A}\ }\textbf {\bibinfo {volume} {3}},\ \bibinfo {pages} {25} (\bibinfo
  {year} {1971})}\BibitemShut {NoStop}%
\end{thebibliography}%

\end{document}